\documentclass[twocolumn]{aastex63}

\usepackage{color}
\usepackage{CJK}
\usepackage{graphicx}
\usepackage{amsmath}
\usepackage{amssymb}
\usepackage{appendix}

\usepackage{lineno}
\definecolor{chmagenta}{rgb}{0.54, 0.17, 0.88}

\graphicspath{{./}{figures/}}

\shorttitle{Short GRB Hosts II: Stellar Populations}
\shortauthors{Nugent et al.}

\begin{document}
\begin{CJK*}{UTF8}{gbsn}

\title{Short GRB Host Galaxies. II. A Legacy Sample of Redshifts, Stellar Population Properties, and Implications for their Neutron Star Merger Origins}

\correspondingauthor{A. E. Nugent}
\email{anyanugent2023@u.northwestern.edu}

\newcommand{\NU}{\affiliation{Center for Interdisciplinary Exploration and Research in Astrophysics (CIERA) and Department of Physics and Astronomy, Northwestern University, Evanston, IL 60208, USA}}

\newcommand{\Purdue}{\affiliation{Purdue University, 
Department of Physics and Astronomy, 525 Northwestern Avenue, West Lafayette, IN 47907, USA}}

\newcommand{\CfA}{\affiliation{Center for Astrophysics\:$|$\:Harvard \& Smithsonian, 60 Garden St. Cambridge, MA 02138, USA}}

\newcommand{\UCSC}{\affiliation{Department of Astronomy and Astrophysics, University of California, Santa Cruz, CA 95064, USA}}

\newcommand{\IS}{\affiliation{Centre for Astrophysics and Cosmology, Science Institute, University of Iceland, Dunhagi 5, 107 Reykjav\'ik, Iceland}}

\newcommand{\DAWN}{\affiliation{Cosmic Dawn Center (DAWN), Niels Bohr Institute, University of Copenhagen, Jagtvej 128, 2100 Copenhagen \O, Denmark}}

\newcommand{\PUCV}{\affiliation{Instituto de F\'isica, Pontificia Universidad Cat\'olica de Valpara\'iso, Casilla 4059, Valpara\'iso, Chile}}

\newcommand{\IPMU}{\affiliation{Kavli Institute for the Physics and Mathematics of the Universe (Kavli IPMU), 5-1-5 Kashiwanoha, Kashiwa, 277-8583, Japan}}

\newcommand{\PSU}{\affiliation{Department of Astronomy \& Astrophysics, The Pennsylvania State University, University Park, PA 16802, USA}}

\newcommand{\ICDS}{\affiliation{Institute for Computational \& Data Sciences, The Pennsylvania State University, University Park, PA, USA}}

\newcommand{\IGC}{\affiliation{Institute for Gravitation and the Cosmos, The Pennsylvania State University, University Park, PA 16802, USA}}

\newcommand{\Swin}{\affiliation{ Centre for Astrophysics and Supercomputing, Swinburne University of Technology, Hawthorn, VIC, 3122, Australia}}

\newcommand{\Curtin}{\affiliation{ International Centre for Radio Astronomy Research, Curtin University, Bentley, WA 6102, Australia}}

\newcommand{\MQ}{\affiliation{Department of Physics \& Astronomy, Macquarie University, NSW 2109, Australia}}

\newcommand{\MQAAAstro}{\affiliation{Macquarie University Research Centre for Astronomy, Astrophysics \& Astrophotonics, Sydney, NSW 2109, Australia}}

\newcommand{\CSIRO}{\affiliation{CSIRO, Space and Astronomy, PO Box 76, Epping NSW 1710 Australia}}

\newcommand{\KICP}{\affiliation{Kavli Institute for Cosmological Physics, The University of Chicago, 5640 South Ellis Avenue, Chicago, IL 60637, USA}}

\newcommand{\UA}{\affiliation{University of Arizona, Steward Observatory, 933~N.~Cherry~Ave., Tucson, AZ 85721, USA}}

\newcommand{\EFI}{\affiliation{Enrico Fermi Institute, The University of Chicago, 933 East 56th Street, Chicago, IL 60637, USA}}

\newcommand{\mpia}{\affiliation{Max-Planck-Institut f\"{u}r Astronomie (MPIA), K\"{o}nigstuhl 17, 69117 Heidelberg, Germany}}

\newcommand{\GWU}{\affiliation{Department of Physics, The George Washington University, Washington, DC 20052, USA}}

\newcommand{\UCB}{\affiliation{Department of Astronomy, University of California, Berkeley, CA 94720-3411, USA}}

\newcommand{\RU}{\affiliation{Department of Astrophysics/IMAPP, Radboud University, PO Box 9010,
6500 GL, The Netherlands}}

\newcommand{\LJMU}{\affiliation{Astrophysics Research Institute, Liverpool John Moores University, IC2, Liverpool Science Park, 146 Brownlow Hill, Liverpool L3 5RF, UK}}

\newcommand{\LU}{\affiliation{School of Physics and Astronomy, University of Leicester, University Road, Leicester. LE1 7RH, UK}}
\author[0000-0002-2028-9329]{Anya E. Nugent}
\NU

\author[0000-0002-7374-935X]{Wen-fai Fong}
\NU

\author[0000-0002-9363-8606]{Yuxin Dong (董雨欣)}
\NU

\author[0000-0001-6755-1315]{Joel Leja}
\PSU
\ICDS
\IGC

\author[0000-0002-9392-9681]{Edo Berger}
\CfA

\author[0000-0002-0147-0835]{Michael Zevin}
\KICP\EFI

\author[0000-0002-7706-5668]{Ryan Chornock}
\UCB

\author[0000-0002-9118-9448]{Bethany E. Cobb}
\GWU

\author[0000-0002-6625-6450]{Luke Zoltan Kelley}
\NU

\author[0000-0002-5740-7747]{Charles D. Kilpatrick}
\NU

\author[0000-0001-7821-9369]{Andrew Levan}
\RU

\author[0000-0003-4768-7586]{Raffaella Margutti}
\UCB

\author[0000-0001-8340-3486]{Kerry Paterson}\mpia

\author[0000-0001-8472-1996]{Daniel Perley}\LJMU

\author[0000-0003-3937-0618]{Alicia Rouco Escorial}
\NU

\author[0000-0001-5510-2424]{Nathan Smith}\UA

\author[0000-0003-3274-6336]{Nial Tanvir}
\LU

\begin{abstract}
We present the stellar population properties of 69 short gamma-ray burst (GRB) host galaxies, representing the largest uniformly-modeled sample to date. Using the \texttt{Prospector} stellar population inference code, we jointly fit photometry and/or spectroscopy of each host galaxy. We find a population median redshift of $z = 0.64^{+0.83}_{-0.32}$ ($68\%$ confidence), including 9 photometric redshifts at $z\gtrsim1$. We further find a median mass-weighted age of $t_m$=$0.8^{+2.71}_{-0.53}$~Gyr, stellar mass of log(M$_*$/M$_\odot$)=$9.69^{+0.75}_{-0.65}$, star formation rate of SFR=$1.44^{+9.37}_{-1.35}$M$_\odot$yr$^{-1}$, stellar metallicity of log(Z$_*$/Z$_\odot$)=$-0.38^{+0.44}_{-0.42}$, and dust attenuation of $A_V=0.43^{+0.85}_{-0.36}$~mag (68\% confidence). Overall, the majority of short GRB hosts are star-forming ($\approx84\%$), with small fractions that are either transitioning ($\approx6\%$) or quiescent ($\approx10 \%$); however, we observe a much larger fraction ($\approx40\%$) of quiescent and transitioning hosts at $z\lesssim0.25$, commensurate with galaxy evolution. We find that short GRB hosts populate the star-forming main sequence of normal field galaxies, but do not include as many high-mass galaxies as the general galaxy population, implying that their binary neutron star (BNS) merger progenitors are dependent on a combination of host star formation and stellar mass. The distribution of ages and redshifts implies a broad delay-time distribution, with a fast-merging channel at $z>1$ and a decreased neutron star binary formation efficiency from high to low redshifts. If short GRB hosts are representative of BNS merger hosts within the horizon of current gravitational wave detectors, these results can inform future searches for electromagnetic counterparts. All of the data and modeling products are available on the BRIGHT website.
\end{abstract}
% SHINE BRIGHT!


\section{Introduction}
\label{sec:intro}
\end{CJK*}
Short-duration gamma-ray bursts (GRBs) are some of the most luminous explosions in the universe and are observed over a large range of cosmological distances ($z \approx 0.1-2.5$; \citealt{skm+18,pfn+20}). Host galaxy associations to short GRBs have been pivotal in uncovering the true nature of GRB progenitors, as they provide redshifts and information on their stellar populations and the types of stars that produced them. A few of the first short GRBs detected soon after the launch of the \textit{Neil Gehrels Swift} Observatory (\textit{Swift}; \citealt{ggg+04}) originated in older, quiescent host galaxies (GRBs 050509B and 050724; \citealt{bpc+05,gso+05, bpp+06}), implying that short GRBs are derived from older stellar progenitors than long-duration GRBs, with an assumed long formation timescale. However, within a few years, studies showed that a majority of short GRBs ($\approx 75\%$) were associated with younger, star-forming galaxies, indicating that their progenitors have a spread of formation timescales \citep{lb10, fb13, ber14}. Coupled with their energy scales \citep{gbb+08}, inferred event rates \citep{fbm+15}, locations within their host galaxies \citep{fb13,tlt+14}, lack of associated supernovae (e.g., \citealt{ffp+05,hwf+05,sbk+06}), and claimed detections of the first $r$-process kilonovae \citep{bfc13,tlf+13}, there is a wealth of circumstantial evidence that short GRBs originate from compact object mergers, specifically either in binary neutron star (BNS) or neutron star black hole (NSBH) mergers. 

The LIGO/Virgo discovery of gravitational wave (GW) BNS merger GW170817 and its associated electromagnetic (EM) counterpart, short GRB\,170817A \citep{aaloc+17, gvb+17, sfk+2017} provided a definitive connection between short GRBs and compact object binary mergers for the first time. The host of GW170817/GRB\,170817A, NGC4993, was the oldest by several gigayears and one of the most quiescent galaxies compared to other short GRB hosts and the general field galaxy population \citep{bbf+17, llt+2017, pht+17}. Due to the older stellar population in this host, the ``delay time'' between the formation of the binary system and the BNS merger was likely several gigayears. The characterization of NGC4993, together with the cosmological population of short GRB host environments, motivates us to ask the following questions:

\begin{itemize}
    \item What is the relationship between short GRB progenitors and recent star formation and stellar mass in galaxies?
    \item What are short GRB formation channels and timescales?
    \item What are the differences between the environments of cosmological short GRBs and those of local universe BNS/NSBH mergers?
\end{itemize}

To fully understand the short GRB progenitor from formation to merger and the environmental factors that are at play, it is imperative to characterize the properties of short GRB host galaxies over cosmic time. For example, the stellar population ages and redshifts of the host galaxies serve as vital constraints on the delay time distribution (DTD), while star formation rates (SFRs) and stellar masses describe the conditions under which BNS/NSBH mergers form. Previous studies determined a peak in redshift at $z \approx 0.5$ \citep{ber14}, with detection of just a few short GRBs existing at $z > 1.0$ \citep{pfn+20}, where one may expect to find more short delay time progenitors. More recently, \citet{otd+2022} suggested, based on photometric redshift estimates, that there is a larger sample of short GRBs at $z > 1.0$. Additional studies have deduced a median stellar population age of $\sim 1$~Gyr, and star formation rates and stellar masses that span a wide range \citep{ber09, lb10, fb13, ber14, nfd+20}, indicating no clear trends with a single galaxy property. This is unlike other explosive transients, which originate from massive stars (e.g., long GRBs, super-luminous supernovae, hereafter SLSNe) and have been shown to occur in low metallicity, low-mass, young, and actively star-forming galaxies \citep{Svensson2010, Perley2013, Wang2014, Vergani2015, pqy+16, Niino2017, hth+18, Schulze2021}. Type Ia SNe, which have older white dwarf progenitors, are found in both star-forming and quiescent galaxies and generally trace the properties of the field galaxy population \citep{ot79, mdp+05, slp+06, Pan2021, Wiseman2021}. 

By comparing short GRB host environments to the general field galaxy population, we can also determine if their progenitors have preferences for specific galaxy traits. For instance, the short GRB host population and its adherence to standard galaxy relations, including the star-forming main sequence (SFMS) and mass-metallicity relation \citep{gcb+05,Speagle2014, Whitaker2014, Leja2021}, can offer insight on whether or not short GRB hosts are unusual. The frequency of short GRB host types is also informative, as we expect $\approx 80 \%$ of detectable field galaxies out to $z=0.3$ to be star-forming \citep{Dressler1980, Martis2016, Leja2021}; thus, adherence or deviation from this fraction tells us how dependent the progenitor is on recent star formation.

A uniform and holistic study of short GRB stellar populations, which takes advantage of nearly two decades of short GRB host identifications, has yet to be conducted in a systematic way. Given the steady flow of well-localized cosmological short GRBs, and the present era of GW discovery, it is timely to conduct a large sample study. Over the past few years, stellar population modeling has also rapidly progressed to include more sophisticated star formation histories, informed statistics, and the ability to jointly fit photometric and spectroscopic data (e.g., \citealt{nfd+20, jlc+2021}).

\begin{figure}
\centering
\includegraphics[width=0.475\textwidth]{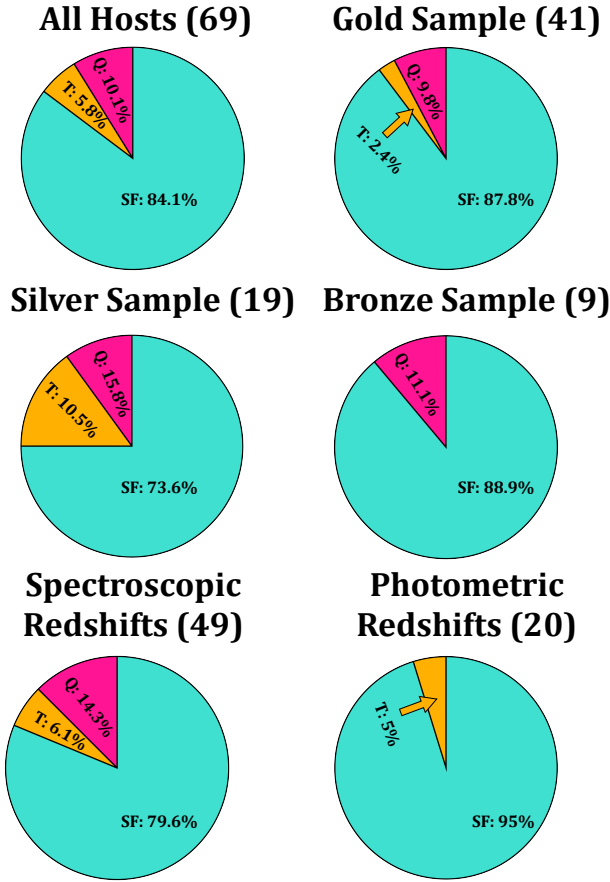}
\vspace{-0.1in}
\caption{The fractions of star-forming (SF; blue), transitioning (T; yellow), and quiescent (Q; red) short GRB host galaxies based on the results in Table \ref{tab:prospectres} for several subgroups of the sample. We see that in the full sample, a majority ($\sim 84\%$) of short GRBs occur in star-forming galaxies, and $\sim 16\%$ occur in transitioning or quiescent galaxies. These fractions also stay fairly consistent for the Gold, Silver, and Bronze samples. We find a lower star-forming fraction for the spectroscopic redshift sample than the photometric redshift sample, likely because the photometric redshift sample reaches higher redshifts, where star-forming galaxies are much more prevalent.}
\label{pie}
\end{figure}

Here, we present the stellar population properties of 69 confidently associated short GRB host galaxies. This represents the second in a series of two papers. Paper I, \citet{BRIGHT-I}, focuses on the photometric and spectroscopic catalogs, host galaxy associations, spectroscopic redshifts, and galactocentric offsets. This paper, Paper II, focuses on spectral energy distribution (SED) modeling of these data, their inferred stellar population properties, and implications for the progenitors. In Section \ref{sec:sample}, we describe the host sample and how hosts were selected for this study. We outline our stellar population modeling methods in Section \ref{sec:methods}. We discuss the inferred stellar population properties and redshift distribution in Section \ref{sec:sp_prop}, and compare short GRB host properties to several known galaxy relations, including the star-forming main sequence, mass-weighted Schechter function, and the mass-metallicity relation. We compare properties intrinsic to the GRB, including their offsets, and their host in Section \ref{sec:grb_prop}. We examine short GRB hosts in the context of other transient hosts and the host of GW170817/GRB\,170817 in Section \ref{sec:sgrb_trans}. We list several possible sample biases and potential missing populations in Section \ref{sec:biases}. Finally, our conclusions are given in Section \ref{sec:conclusions}. We house all of the data and modeling products described in this work on the Broadband Repository for Investigating Gamma-ray burst Host Traits (BRIGHT) website\footnote{\url{http://bright.ciera.northwestern.edu}}.

Unless otherwise stated, all observations are reported in the AB magnitude system and have been corrected for Galactic extinction in the direction of the GRB \citep{MilkyWay,sf11}.  We employ a standard WMAP9 cosmology of $H_{0}$ = 69.6~km~s$^{-1}$~Mpc$^{-1}$, $\Omega_\textrm{M}$ = 0.286, $\Omega_\textrm{vac}$ = 0.714 \citep{Hinshaw2013, blw+14}. 

\begingroup
\setlength{\tabcolsep}{6pt} % Default value: 6pt
\renewcommand{\arraystretch}{1.3} % Default value: 1
\begin{deluxetable*}{l|ccccccc}[!t]
\tabletypesize{\normalsize}
\tablecolumns{3}
\tablewidth{0pc}
\tablecaption{\texttt{Prospector} Parameters and Prior Distributions
\label{prospector_priors}}
\tablehead{
\colhead {Parameter}	 &
\colhead {Definition}	 &
\colhead {Prior}	
}
\startdata
\multicolumn{3}{c}{FITTING PARAMETERS} \\
\hline
$z$ & redshift & $\mathcal{U}[0.1,3.0] $ \\
$t_\textrm{age}$ (Gyr) & age of the galaxy at the time of observation & $\mathcal{U}[0.1,t_\textrm{lookback}(z)]$ \\
$\tau$ & e-folding time of delayed-$\tau$ SFH & $\mathcal{U}[0.1,10.0] $ \\
$\log(M_F/M_\odot)$ & total mass formed & $\mathcal{U}[7,13] $ \\
$\log(Z_*/Z_\odot)$ & stellar metallicity & $\mathcal{U}[-1.0,0.2] $ \\
$\tau_{V,1}$ & optical depth of young stellar light & $\mathcal{U}[0.1,0.5*\tau_{V,2}]$ \\
$\tau_{V,2}$ & optical depth of old stellar light & $\mathcal{U}[0,3] $ \\
$N_0$ & spectral normalization factor & $\mathcal{N}(\mu = 1.0, \sigma=0.2)$ \\
$\log(Z_{gas}/Z_\odot)$ & gas-phase metallicity & $\mathcal{U}[-2.0,0.5] $ \\
$U_{gas}$ & gas ionization parameter & $\mathcal{U}[-4,-1] $ \\
$\tau_{\textrm{AGN}}$ & mid-IR optical depth & $\mathcal{U}[10,90] $ \\
$f_{\textrm{AGN}}$ & fraction of AGN luminosity in galaxy & $\mathcal{U}[10^{-5},2$] \\
$N_{\textrm{jitter}}$ & noise inflation factor for spectra & $\mathcal{U}[1,3] $ \\
\hline
\multicolumn{3}{c}{DERIVED PARAMETERS} \\
\hline
$t_m$ & mass-weighted age in Gyr &  $t_m$ = $t_{\text{age}} - \frac{\int_0^{t_{\text{age}}} t \times \text{SFR}(t) dt}{ \int_0^ {t_{\text{age}}} \text{SFR}(t) dt}$ \\
$\log(M_*/M_\odot)$ & stellar mass &  M$_* \approx$ M$_F \times 10^{1.06 - 0.24\log{(t_m)} + 0.01*\log{(t_m)}^2}$ \\
SFR & star formation rate in M$_\odot$ yr$^{-1}$ & SFR(t) = M$_F\times \Big[\int_0^t{te^{-t/\tau} dt}\Big]^{-1} \times te^{-t/\tau}$ \\
$A_V$ & total dust attenuation in mag & $A_V = 1.086\times(\tau_{V,1} + \tau_{V,2})$
\enddata
\tablecomments{A list of all possible free parameters used in our \texttt{Prospector} fits, their definitions, and their prior distributions. $\mathcal{U}$ represents a uniform distribution, while $\mathcal{N}$ represents a normal distribution. We also list the derived parameters used in this analysis. We note that $z$ is only set free for hosts that do not have a known redshift, $N_0$, $\log(Z_{gas}/Z_\odot)$, and $U_{gas}$ are only used in fits including spectroscopy, the AGN parameters were only included for one host (GRB 150101B) that has a known active galactic nucleus (AGN), and $N_{\textrm{jitter}}$ was only applied to hosts with high signal-to-noise ratio (S/N) spectra.}
\end{deluxetable*}

\section{Short GRB Host Sample}
\label{sec:sample}
\begin{figure*}[t]
\vspace{-0.2in}
\centering{\includegraphics[width=0.50\textwidth]{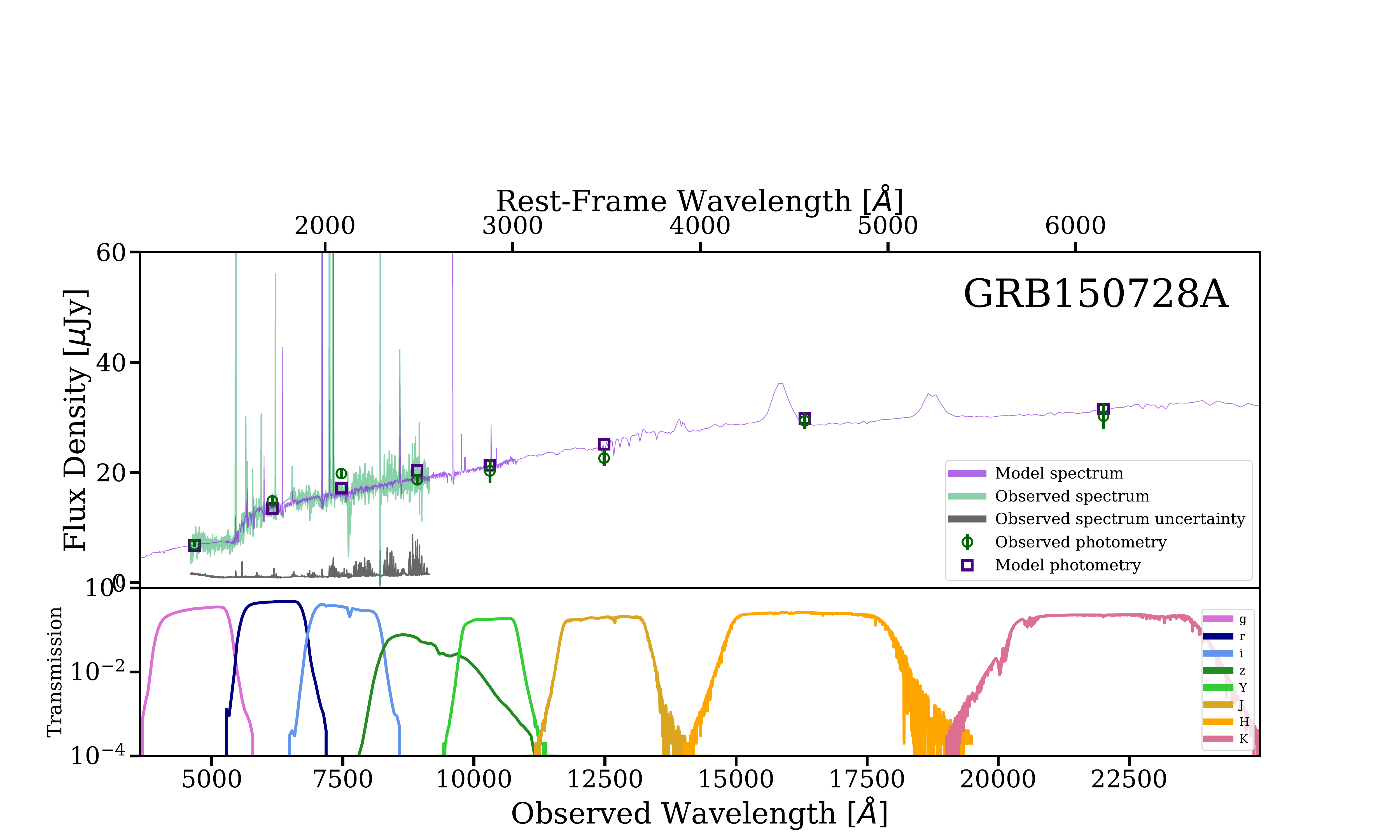}
\hspace{-0.2in}
\includegraphics[width=0.50\textwidth]{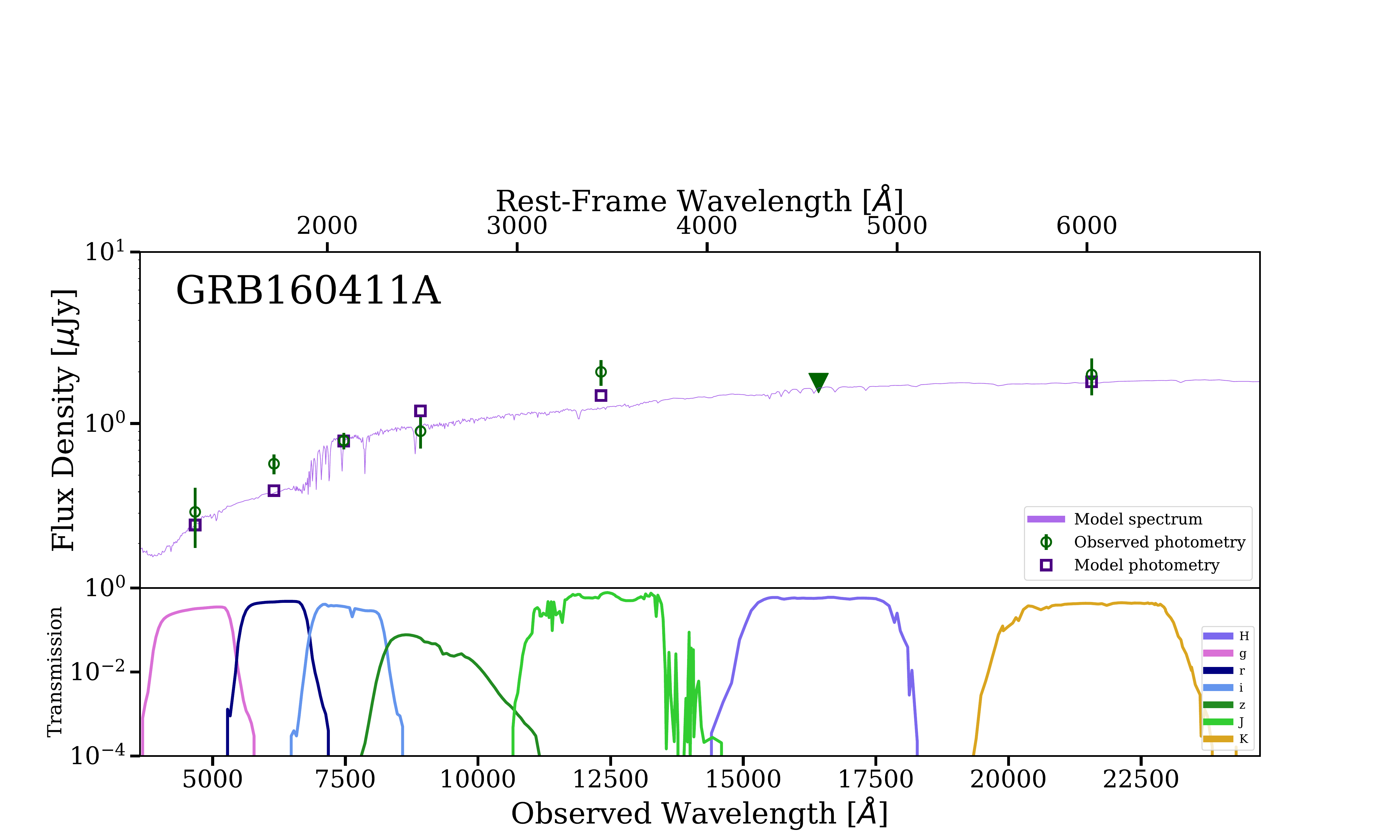}}
\caption{Representative fits from two of the 69 short GRBs in our sample. \texttt{Prospector} produced model spectra (purple lines) and photometry (purple squares) compared to the observed spectra (green line) and error (grey line) and photometry (green circles) for the joint spectroscopy and photometry fit of GRB 150728A (left panel) and the redshift-free, photometry only fit of GRB 160411A (right panel). We highlight these fits to show the accuracy in measurement of the spectral continuum and line locations (GRB 150728A) and the photometric colors for redshift determination (GRB 160411A). In total, we have performed 40 joint spectroscopic and photometric fits and 29 photometry-only fits. These modeling products are all available on the BRIGHT website.}
\label{SED}
\end{figure*}

We use the short GRB host sample described in \citet{BRIGHT-I}\footnote{We note that this sample also includes the long-duration GRBs 060614 and 211211A as there is significant evidence that they are derived from neutron star (NS) mergers and not young, stellar progenitors.}. This sample contains 84 short GRB hosts with broad-band photometric detections and upper limits; $58\%$ of the sample has spectroscopic redshifts and $50\%$ of the sample has available spectra with spectral line detections. Host associations in this sample are defined through the probability of chance coincidence method ($P_\textrm{cc}$; \citealt{bkd02}), which uses an optical magnitude of the host compared to the distance from the host to the short GRB X-ray, optical, or radio afterglow position. A low $P_\textrm{cc}$ translates to a higher likelihood that the short GRB is associated with the galaxy. For this study, we only consider hosts where $P_\textrm{cc} \lesssim 0.20$, above which we cannot conclusively assign a host galaxy to a burst. Furthermore, as the goal of this study is to model the SEDs of the host galaxies, we only use the hosts that are detected in $\geq 3$ photometric bands. Thus, our sample here comprises 69 out of the 84 host associations found in \citet{BRIGHT-I}. 

From here on, we define the ``Gold Sample" as that with the most robust host associations ($P_\textrm{cc} < 0.02$). Similarly, the ``Silver Sample"  ($0.02 < P_\textrm{cc} < 0.09$) and the ``Bronze Sample" ($0.09 < P_\textrm{cc} < 0.20$) represent moderately robust associations (see \citealt{BRIGHT-I} for a more detailed explanation). We also classify hosts based on whether they have photometric (20 hosts) or spectroscopic redshifts (49 hosts). There are 40 hosts (31 with spectroscopic redshifts) in the Gold Sample, 19 hosts (11 with spectroscopic redshifts) in the Silver Sample, and 9 (7 with spectroscopic redshifts) hosts in the Bronze Sample. We show this breakdown in Figure \ref{pie}. 

\section{Stellar population Modeling}
\label{sec:methods}
To determine the stellar population properties of the host galaxies, we use the Python-based SED modeling code, \texttt{Prospector} \citep{Leja_2017, jlc+2021}. We jointly fit photometric and spectroscopic data (when available) to determine properties including redshift (when not already determined from the spectrum), stellar mass, stellar population age, gas and stellar metallicities, and dust attenuation (see Table \ref{prospector_priors} for a review of fit properties). We apply the nested sampling fitting routine \texttt{dynesty} \citep{Dynesty} within \texttt{Prospector} to all available observational data to produce posterior distributions of the sampled properties. \texttt{Prospector} applies the \texttt{MIST} \citep{MIST} models and \texttt{MILES} spectral library \citep{MILES} through \texttt{FSPS} (Flexible Stellar Population Synthesis) and \texttt{python-FSPS} to produce model photometry and spectra \citep{FSPS_2009, FSPS_2010}.

For all \texttt{Prospector} fits, we use a Chabrier initial mass function (IMF; \citealt{Chabrier2003}), Milky-Way extinction law \citep{MilkyWay}, a parametric delayed-$\tau$ star formation history (SFH; $\text{SFH} \propto t*e^{-t/\tau}$) defined by the e-folding factor $\tau$, and include the effects of nebular emission and line strengths through a nebular emission model \citep{bdc+2017}. We choose a Milky Way extinction law as it has been shown to model the majority of galaxy SED shapes well, and moreover the majority of the hosts in this sample. We also include the \citet{gcb+05} mass-metallicity relation of galaxies in the fits, which ensures realistic mass values are being sampled for a given sampled metallicity. For star-forming hosts, we determine the dust attenuation from old ($\geq 10^7$~yr; \citealt{FSPS_2010}) and young stars, respectively, through the commonly-used 2:1 ratio \citep{pkb+14,cab+20, Leja2019}, as younger stars tend to attenuate twice the amount of dust as old stars. For known quiescent hosts, we assume there is no dust attenuation of young stellar light (parameter $\tau_{V,1}=0$). 

The observed host photometry is modeled in \texttt{Prospector} by integrating over the wavelength coverage of the filter transmission curves. For most host galaxies, we use the standard Sloan Digitial Sky Survey (SDSS; $griz$; \citealt{SDSSFilters}), Bessell ($BVRI$), Two Micron All Sky Survey (2MASS; $JHK$; \citealt{2Mass}), Spitzer, Wide-field Infrared Survey Explorer (WISE; \citealt{WISE}), and Wide Field Camera 3 WFC3/IR and WFC3/UVIS (\citealt{WFC3}) transmission curves, available in the \texttt{SedPy} Python package (10.5281/zenodo.4582722). We apply telescope-specific filter transmission curves when photometry is from the Keck Observatory (LRIS, DEIMOS, MOSFIRE), the MMT Observatory (MMIRS), and UKIRT (WFCAM), as these differ from the standard filter sets. The instruments used for the photometric data of the hosts are reported in Table~1 in \citet{BRIGHT-I}.

For all hosts, the main fitted parameters are redshift ($z$) if it is not known, mass formed ($M_F$), age of the galaxy at the time of observation ($t_\textrm{age}$), the SFH e-folding factor $\tau$, optical depth due to dust of young ($\tau_{V,1}$) and old ($\tau_{V,2}$) stellar light, and stellar metallicity ($Z_*$) (see Table \ref{prospector_priors}). We perform three main types of fits depending on the data available, each of which contain unique parameter specifications and sometimes added stellar population properties: (i) joint photometric and spectroscopic fits with known redshifts (see Figure \ref{SED}, left panel); (ii) photometric fits with known redshifts; and (iii) photometric fits with no known redshifts (see Figure \ref{SED}, right panel).

First, we perform joint spectroscopic and photometric \texttt{Prospector} fits for all hosts with at least one spectral line detection of signal-to-noise ratio (S/N) $\geq 5$ (we find diminishing returns for host galaxies with lower S/N). In our sample, we have 40 hosts with spectra that meet these conditions (for an example, see Figure \ref{SED}, left panel). We determine redshifts of these hosts from spectral line detection (see \citealt{BRIGHT-I}) and fix the $z$ parameter to the spectroscopic redshift and the maximum of $t_\textrm{age}$ to the lookback time at that redshift ($t_\textrm{lookback}(z)$). We include additional free parameters that are dependent on spectral line detection in these fits to better estimate the observed spectral line strengths and continuum: gas-phase metallicity ($Z_\textrm{gas}$), the dimensionless gas ionization parameter that measures the ratio of hydrogen ionizing photons density to hydrogen density ($U_\textrm{gas}$), and a parameter which normalizes the model to the observed continuum ($N_0$). The $N_0$ parameter is also used to marginalize over flux calibration uncertainties and slit losses. The model spectral continuum is built from an $n^{th}$-order Chebyshev polynomial\footnote{We typically set $n = 10$, although we increase it to 12 for spectra that cover a wider observed wavelength range ($\geq 4000$\AA) and have larger fluctuations in the observed continua, and decrease it to $n=6$ for spectra that have a smaller wavelength coverage ($\leq 3000$\AA) or very flat continua. We note that using a higher $n$ than necessary overfits the continuum, which can affect true spectral line strength.}. In addition, for some hosts we apply a spectral noise inflation model to properly weight the photometric observations against the high S/N spectrum\footnote{These include GRBs 050724, 051221A, 061006, 101224A, 120305A, 140903A, 150728A, 170428A, and 201221D.}. The host of GRB 150101B contains a known active galactic nucleus (AGN; \citealt{fmc+16, xft+2016}). Thus we also include several AGN parameters to properly account for the optical depth in the mid-IR ($t_\textrm{AGN}$) and fraction of the AGN luminosity in the galaxy ($f_\textrm{AGN}$). 

\begin{figure*}[t]
\makebox[\textwidth][c]{\includegraphics[width=1.0\textwidth]{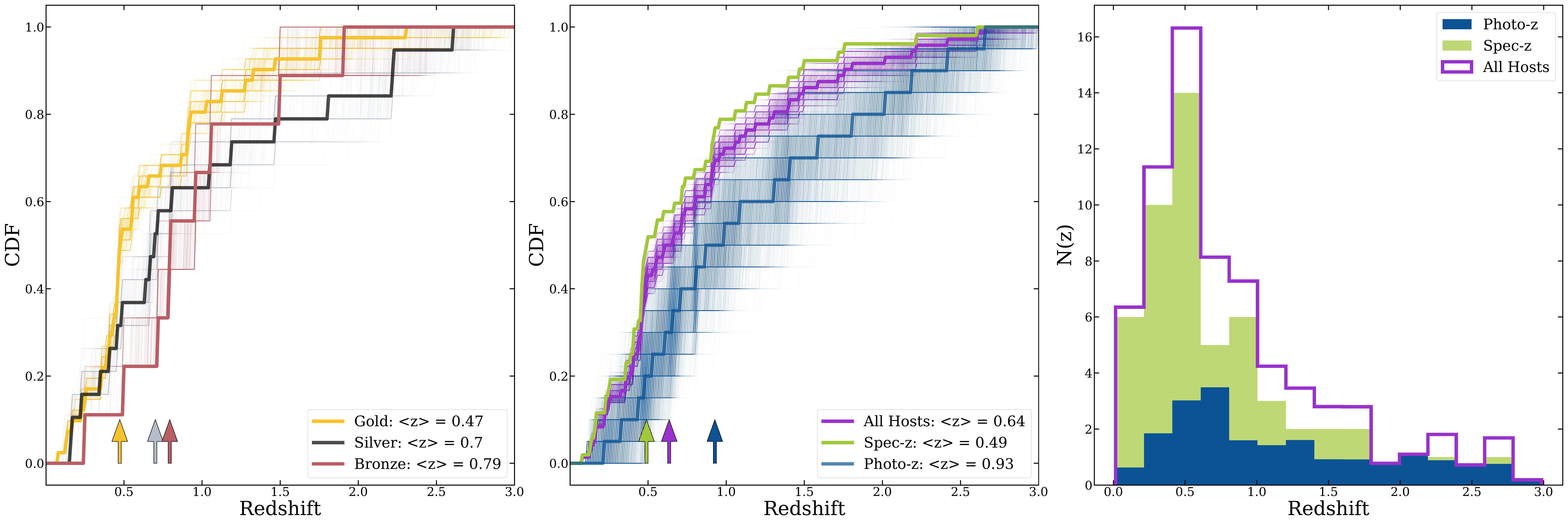}}
\vspace{-0.1in}
\caption{{\bf Left:} The cumulative distribution function (CDF) in redshift for the Gold, Silver, and Bronze samples, as well as 5000 realizations of the CDF based on random draws from the individual posterior distributions. The thicker lines represent the median CDF while the thin lines show the individual realizations to demonstrate the sample spread. We find that there is an increasing redshift median from the Gold to Bronze samples, as indicated by the arrows. {\bf Middle:} The CDF and 5000 realizations in redshift of all short GRB hosts in our sample (purple), along with the photometric redshift (blue) and spectroscopic redshift sample (green). {\bf Right:} Redshift histograms of all hosts (purple), spectroscopic redshift hosts (green), and photometric redshift hosts (blue). The ``All" hosts and photometric redshift distributions are built from the individual posterior distributions as described in Section \ref{sec:sp_overview}. The photometric sample hosts most notably fill in the ``redshift desert" ($z > 1.0$), and the photometric redshift sample has a median $\sim 0.5$ greater than the spectroscopic redshift sample.}
\label{redshift}
\end{figure*}

The second type of fit is for short GRB hosts with known spectroscopic redshifts, but no available or very low S/N spectra for the \texttt{Prospector} fits. For a majority of these hosts, we found spectroscopic redshifts in previously published works, but could not find publicly available spectra. For both the hosts of GRBs 140930B and 181123B \citep{pfn+20}, their spectral continua have low S/N and are not meaningful in the \texttt{Prospector} fits. Thus, we perform a photometry-only fit for these hosts, where we set the $z$ and maximum $t_\textrm{age}$ parameters in the same manner as the joint spectroscopic and photometric fits. Furthermore, we fix $Z_\textrm{gas} = Z_\odot$ and $U_\textrm{gas} = -2.0$ and remove nebular emission lines in the model spectra, as there are no spectral lines to fit \citep{Leja_2017}. 

The third type of fit is for the 20 hosts with unknown redshifts; in these cases, spectroscopy is not feasible due to the apparent faintness of the hosts or there were no spectral lines detected in their observed spectrum (Figure \ref{SED}, right panel). We therefore leave the redshift as a free parameter and allow it to range uniformly between $0.1 \leq z \leq 3.0$. We choose $z=3.0$ as the maximum possible redshift as \textit{Swift}'s sensitivity to short GRBs steeply drops beyond $z\approx 1$, likely due to detectability of the bursts' luminosity \citep{lsb+16}. Additionally, we set the maximum of $t_\textrm{age}$ to be the maximum age of the universe at the sampled redshift. For the hosts of GRBs\,130515A, 160411A, and 180418A, we see large increases in flux between two photometric bands ($r$ and $i$, $z$ and $J$, and $z$ and $J$, respectively), which is a clear indication of the $4000$ \AA\ break. Therefore, we tighten their redshift ranges to only allow for redshifts that give a $4000$ \AA\ break within those wavelength ranges. For GRB\,180418A, we have further knowledge of its possible redshift ($z > 1.0$) from \citet{rfv+2021}. For the host of GRB 210726A, we restrict the redshift range to $z < 1.0$, as solutions at $z > 1.0$ violate a deep $U$-band upper limit and its afterglow luminosity suggests a less likely $z > 1.0$ origin (Schroeder et al. in prep.). For the hosts of GRBs\,170127B and 210726A, we use the upper limits in $J$ and $K$ bands and $U$ and $J$ bands, respectively, in the fits to better constrain photometric redshift estimate.

\begin{figure}
\centering
\includegraphics[width=0.35\textwidth]{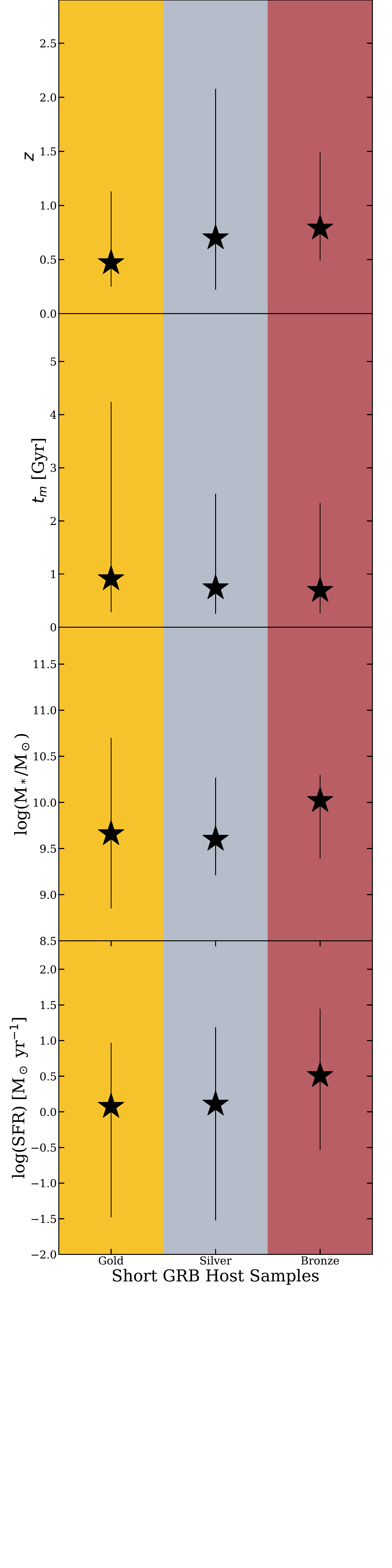}
\vspace{-1.8in}
\caption{\textit{From top to bottom:} median (star) and 8$\%$ confidence interval values (line) for $z$, $t_m$, $\log(M_*/M_\odot)$, and SFR for the Gold, Silver, and Bronze Samples. There is little difference in the median and 68$\%$ confidence interval for the properties in each sample; however, redshift does appear to increase from the Gold to Bronze Samples. \citet{BRIGHT-I} also showed that there are weaker associations to hosts as redshift increases, due to faintness of the galaxies.}
\label{gsb}
\end{figure}

As a point of comparison, we also fit the photometric data of NGC4993 (the host of GRB\,170817A). We fit the GALEX, PS1, 2MASS, and WISE photometric data (applying the respective filter transmission curves) and redshift of NGC4993 listed in \citet{bbf+17}, in which a nonparametric SFH \texttt{Prospector} model fit was achieved. Parametric fits and results are preferred for this work, as they better establish uniformity amongst datasets with inconsistent amount and quality of data. We also note that typically, nonparametric fits result in mass values that are 25-100\% larger and age values that are three to five times older \citep{Leja2019}. Although a {\tt Prospector} fit with a nonparametric SFH was performed in \citet{bbf+17}, for consistency, we perform a parametric SFH fit using the specifications for the photometry and known redshift fits. We do not include NGC4993 as a host in our catalog as the GRB was discovered and associated in a different way than the rest of the host population. 

For a direct comparison to commonly-used physical properties of galaxies, we derive a number of properties from our fitting parameters (Table \ref{prospector_priors}). For instance, we convert mass formed ($M_F$) to stellar mass ($M_*$) using Equation 3 in \citet{nfd+20} and referenced in Table \ref{prospector_priors}. We also use the posteriors on $t_\textrm{age}$ and $\tau$ to derive the posterior on the mass-weighted age ($t_m$). Mass-weighted ages importantly do not overestimate the contribution from young stars, a major caveat of light-weighted or single stellar population (SSP) ages \citep{Conroy2013SED}. We derive the present-day SFR from the posteriors on $t_\textrm{age}$, $\tau$, and $M_F$. Finally, we derive the total dust attenuation $A_V$ from the optical depths: $\tau_{V,1}$ and $\tau_{V,2}$.

\begin{figure*}[t]
\advance\rightskip-0.7in
\hspace{-0.5in}
\includegraphics[width=0.4\textwidth]{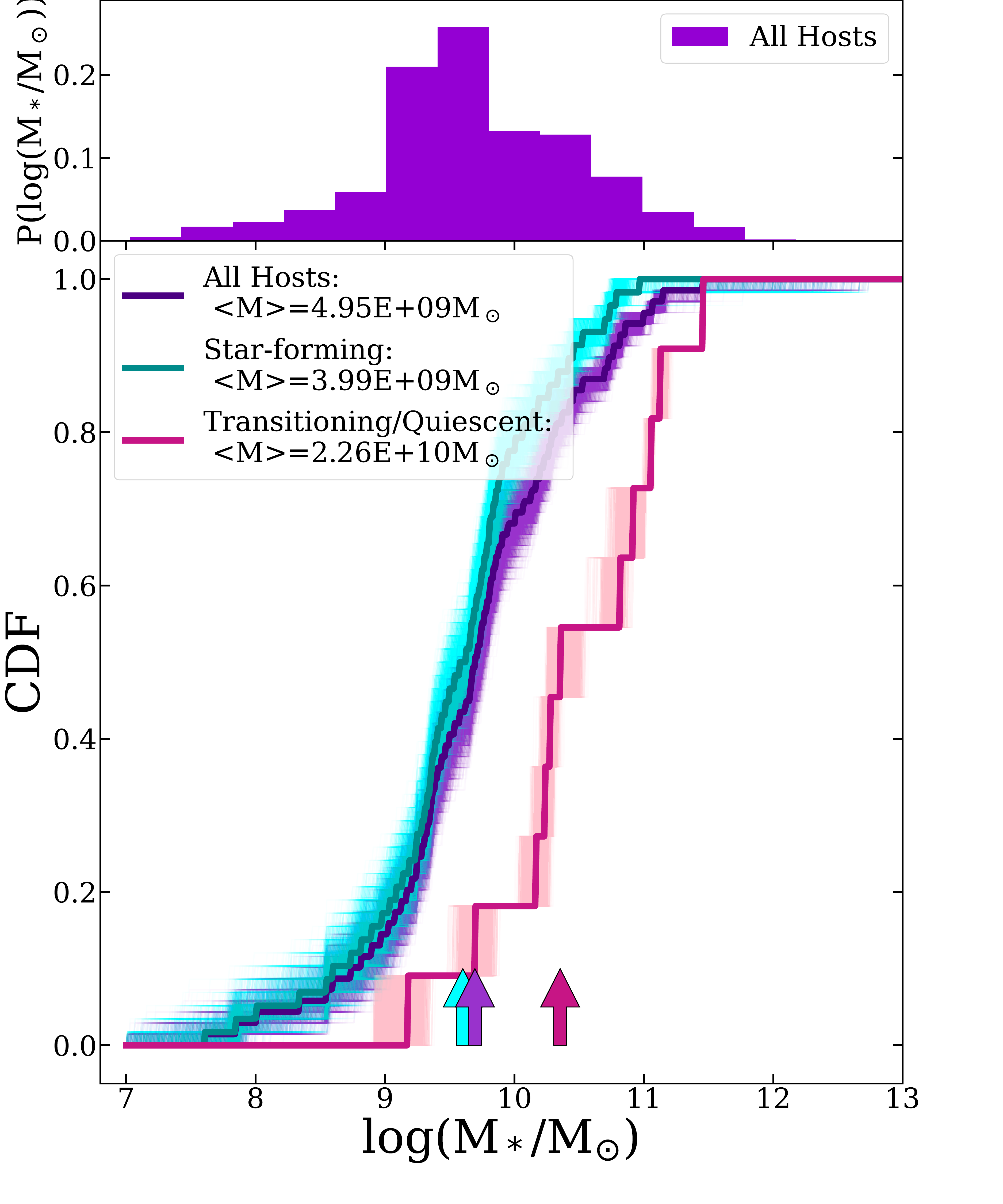}
\hspace{-0.4in}
\includegraphics[width=0.4\textwidth]{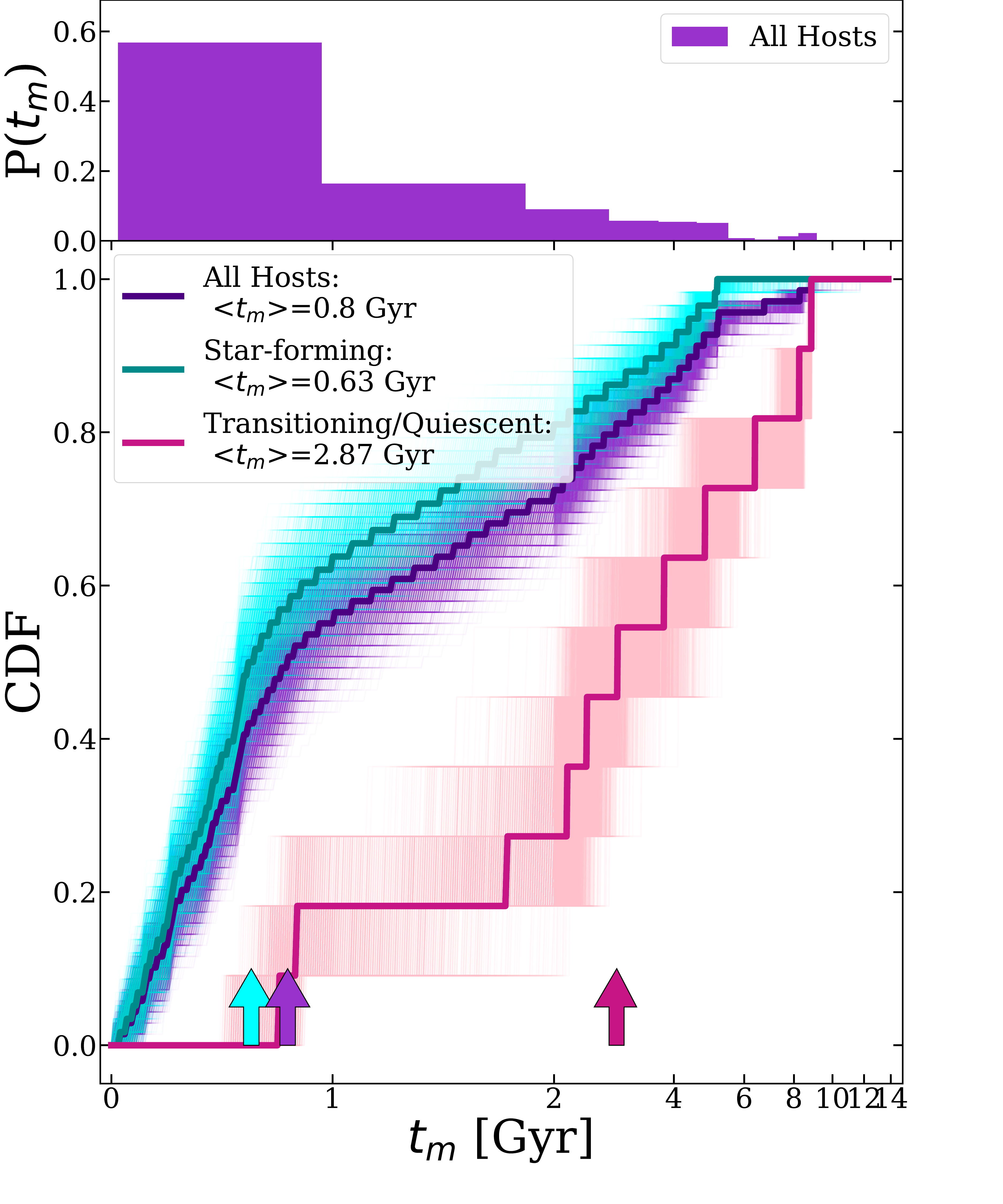}
\hspace{-0.4in}
\includegraphics[width=0.4\textwidth]{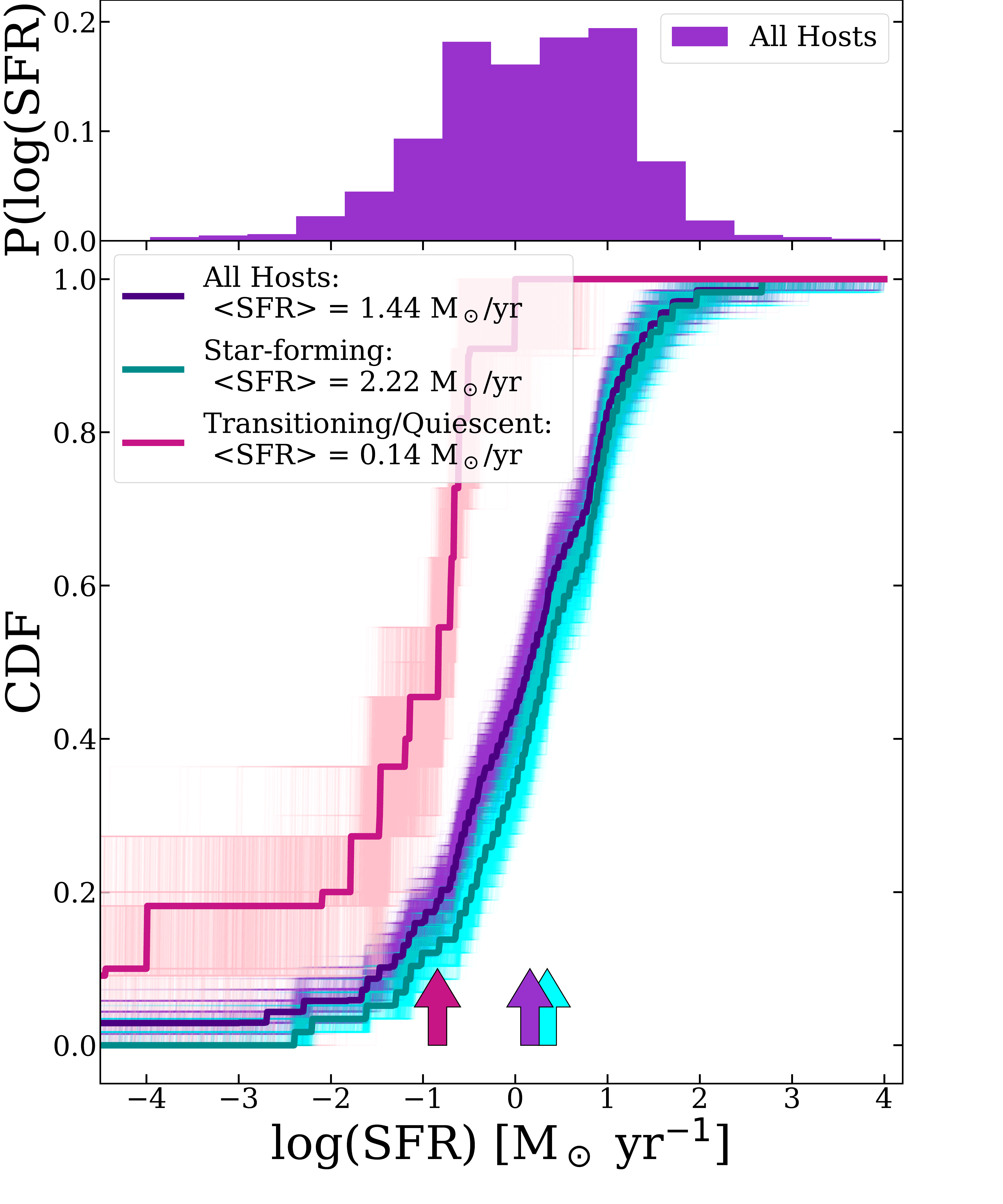}
\includegraphics[width=0.4\textwidth]{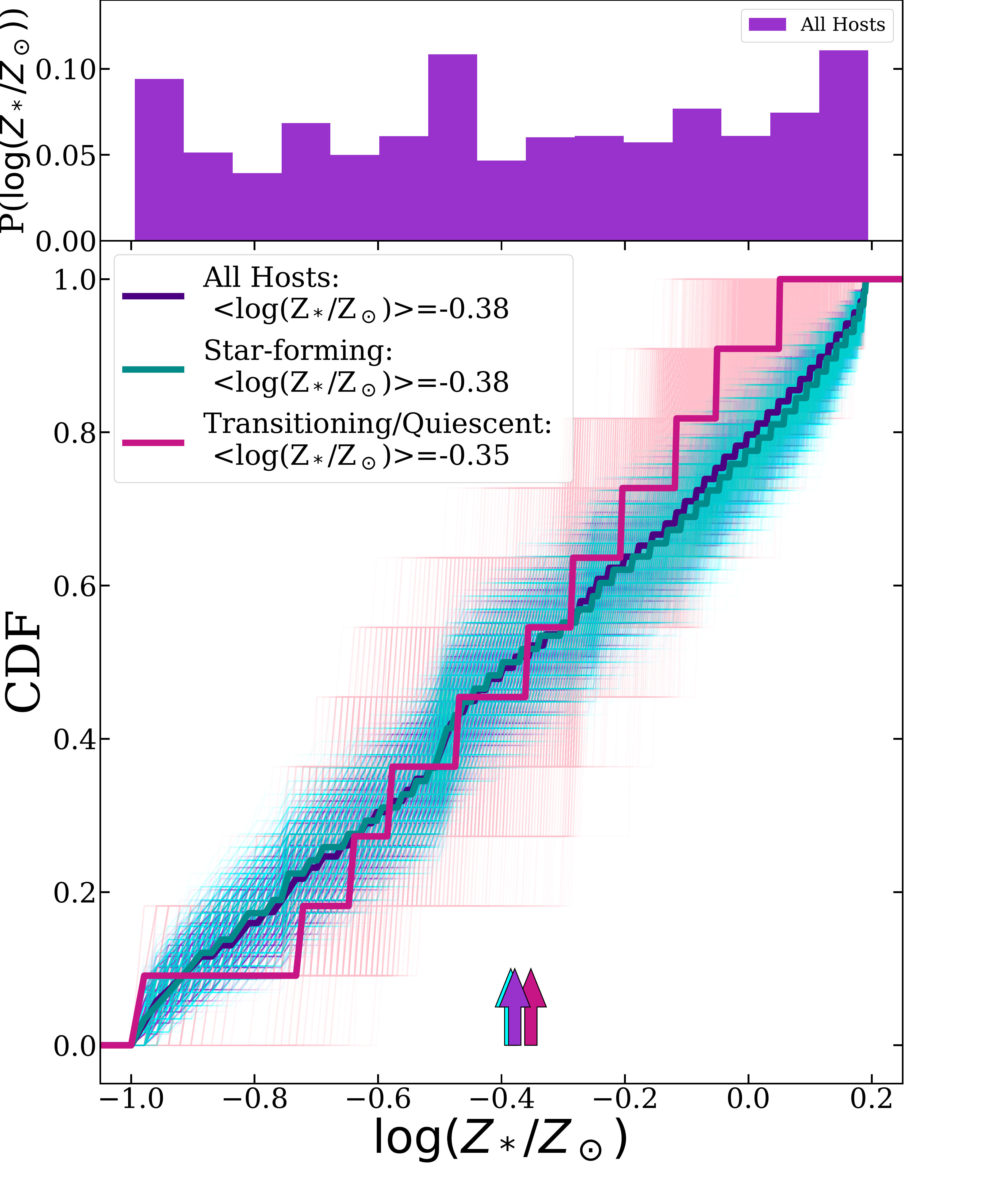}
\includegraphics[width=0.4\textwidth]{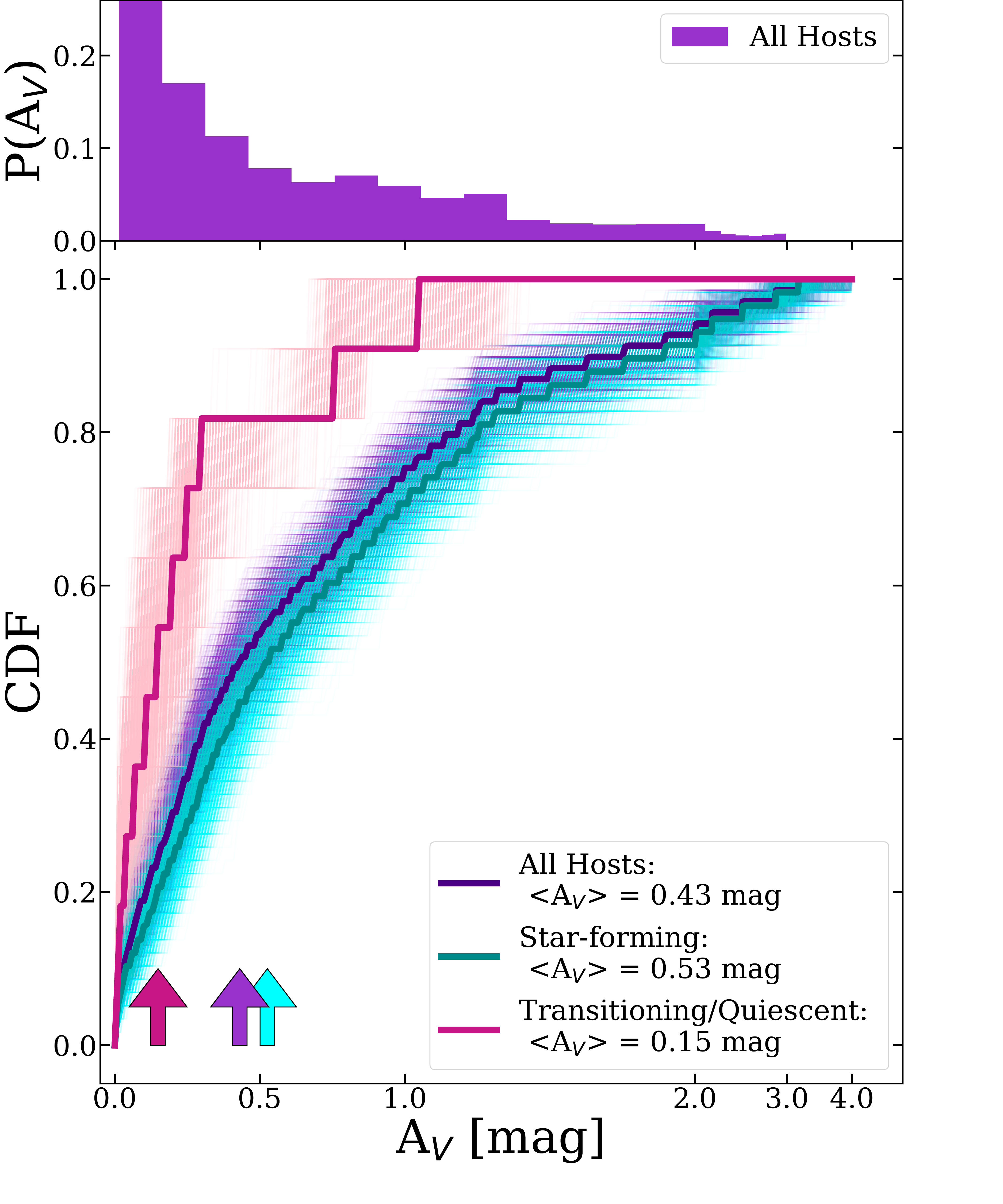}
\caption{\textbf{From left to right, top to bottom:} CDFs of all (purple), star-forming (blue), and transitioning and quiescent (pink) short GRB hosts for log(M$_*$/M$_\odot$), $t_m$ in Gyr, log(SFR) in M$_\odot$ yr$^{-1}$, log(Z$_*$/Z$_\odot$), and $A_V$ in mag. Thick solid lines represent medians on the CDF, while 5000 realizations of the CDF in each distribution are plotted as thin lines to demonstrate the spread in each distribution. The probability distributions of all hosts are shown in the panel above the CDFs. Medians in each distribution are denoted by color-coded arrows from the bottom, and are also denoted in the legend. We find that the median short GRB hosts tend to be massive ($\sim 10^9-10^{10}$M$_\odot$), moderately old ($\sim 0.8$~Gyr), albeit with large difference in ages between the star-forming and transitioning/quiescent hosts, have moderately low ongoing SFR ($\sim 1.5$ M$_\odot$ yr$^{-1}$), have a uniform distribution of stellar metallicities, and have low dust attenuation ($\sim 0.4$~mag).}
\label{cdf_pdf}
\end{figure*}

We list all possible sampled properties, their prior distributions, and the derived parameters $M_*$, $t_m$, SFR, and $A_V$ in Table \ref{prospector_priors}. For each host, we report the median and $68\%$ credible interval of the posterior in several relevant stellar population properties in Table \ref{tab:prospectres}. We report upper limits when the $99.7\%$ credible interval is consistent with the lower limit of the property range.

\section{Stellar Population Properties}
\label{sec:sp_prop}
\subsection{Redshift Distributions}
\label{sec:sp_overview}
Here, we discuss the results of our {\tt Prospector} fitting of 69 short GRB host galaxies (Tables~\ref{tab:prospectres}-\ref{tab:allgroup}). We show the cumulative and posterior distributions in redshift in Figure \ref{redshift}. We report 20 photometric redshifts, 10 of which are completely new results (we show the photometric redshift SEDs in Appendix \ref{sec:photoz}). We build the cumulative distributions (CDF) for the entire host sample from 5000 random draws from individual host \texttt{Prospector}-derived posterior distributions. We include the redshifts of GRBs 090426A ($z=2.609$), 160410A ($z = 1.717$), and 150423A ($z=1.394$) in our sample as spectroscopic redshifts, as all bursts have known redshifts from their afterglows \citep{adp+09, fbc+13, smg+19, atk+2021}. To build the CDF and account for the individual measurement uncertainties, including both spectroscopic and photometric redshifts, we perform 5000 draws of their respective redshift values (for spectroscopic redshifts) or from the {\tt Prospector} posteriors (for photometric redshifts). We choose 5000 draws as there is no change in the results if we increase the number draws beyond 5000. From the 5000 CDFs, we then determine the median CDF from the median value of the realizations in each redshift bin. We find that the spectroscopic redshifts range over $0.1 \leq z \leq 2.2$, with a median of $z = 0.47^{+0.58}_{-0.25}$ (68\% confidence interval; Figure \ref{redshift}) . The addition of the photometric redshift extends the redshift range to $z = 2.8$, and increases the median to $z = 0.64$. This is not surprising, as the photometric redshifts tend to capture $z > 1.0$ (with a median $z \sim 0.93$). This is the so-called ``redshift desert", a region for which it is difficult to obtain spectroscopic redshifts given that the primary identifiable spectral lines are shifted to infrared wavelengths, and these more distant sources are also apparently faint.

Given that 10 of our photometric redshifts are at $z>1$, we also want to ensure that the photometric redshifts are not simply sampling the redshift prior (see Table \ref{prospector_priors}), which could skew them toward higher redshifts. We compare all photometric redshift posterior distributions to their respective prior distributions (which are uniform in linear space) and run an Anderson-Darling (AD) test, with the null hypothesis that both are derived from the same underlying distribution. The AD test best captures the overall shape of the distributions, including any tails. If the p-value of the null hypothesis $P_{AD} < 0.05$, we can reject the null hypothesis. We find that for all short GRB hosts with photometric redshifts, the $P_{AD} \lesssim 0.001$, implying the posterior distributions are not derived from their uniform priors. Thus, the fact that the photometric redshifts are higher than the spectroscopic sample is a real effect and not simply a product of the more limited data. We also ran redshift-free fits for several to test the photometric redshift capabilities of \texttt{Prospector}. For the host of GRB 140930B (which had a very low S/N detection of [O II]), we find a photometric redshift of $z=1.38^{+0.91}_{-0.99}$, which is only 0.09 less than its spectroscopic redshift, although with large error. For the host of GRB 150831A, we find a photometric redshift of $z=1.09^{+0.10}_{-0.19}$, which is 0.09 lower than the true redshift, but within error. Thus, we infer that \texttt{Prospector} generally finds a photometric redshift of the host within its 68\% credible region of the distribution, and is able to capture the low or high redshift nature of the host. Our analysis also highlights the importance of including both spectroscopic and photometric redshifts in the full redshift distribution of short GRB hosts. We also find an increasing median redshift between the Gold, Silver, and Bronze samples, as shown in Figures \ref{redshift} and \ref{gsb}, although they are consistent within the 68\% confidence intervals. This is likely caused by the fact that brighter, lower-redshift short GRBs are easier to associate to host galaxies than much fainter, higher redshift galaxies, as the $P_\textrm{cc}$ test, in part, depends on the brightness of the host (for more, see \citealt{BRIGHT-I}). 

In Figure~\ref{cdf_pdf}, we show the cumulative and posterior distributions for several stellar population properties. The CDFs are derived in the same manner as the redshift distributions. For the entire host sample, we find median stellar population property values of log(M$_*$/M$_{\odot}$) = $9.69^{+0.75}_{-0.65}$, $t_m = 0.8^{+2.71}_{-0.53}$~Gyr, SFR = $1.44^{+9.37}_{-1.35}$ M$_\odot$ yr$^{-1}$, $\log(Z_*/Z_\odot) = -0.38^{+0.44}_{-0.42}$, and $A_V = 0.43^{+0.85}_{-0.36}$~mag. Notably, we find that these properties do not significantly change amongst the Gold, Silver, and Bronze Samples (Figure \ref{gsb}), demonstrating that they are fairly impervious to the confidence in host association. We note that the Gold Sample spans a broader range in mass-weighted age, while the Bronze sample has a slightly elevated SFR (Figure \ref{gsb}); these can naturally be explained as manifestations of the aforementioned redshift differences between the samples.

\subsection{Star Formation Classification}
\label{sec:sf_class}
In order to put the short GRB host sample in context with a basic classification scheme, we systematically classify each host by the degree of star formation. We use the canonical galaxy classifications: ``star-forming'' comprises galaxies on the SFMS, ``transitioning'' comprises galaxies transitioning off the SFMS, and ``quiescent'', comprises galaxies off the SFMS. For this classification, we use the specific SFR (sSFR = SFR/$M_*$ in units of yr$^{-1}$) and the redshift. We determine this classification using Equation 2 in \cite{Tachella2021}: $\mathcal{D}(z) = \text{sSFR} \times t_\text{H}(z)$, where $t_\text{H}(z)$ is the Hubble time at redshift $z$. For hosts with photometric redshifts, we find $\mathcal{D}(z)$ at every sampled $z$ and sSFR. We use the following divisions for classification, following the methods of \cite{Tachella2021}: $\mathcal{D}(z) > 1/3$ describes star-forming galaxies, $1/20 > \mathcal{D}(z) > 1/3 $ defines the galaxies transitioning from star-forming to quiescent, and $\mathcal{D}(z) < 1/20$ represents quiescent galaxies. We define the hosts using the mode of the classification for their distribution of $\mathcal{D}(z)$ values.

In Figure \ref{pie}, we present the percentage of galaxies in each classification in the entire host sample, as well as the Gold, Silver, Bronze, spectroscopic, and photometric redshift samples. The majority of hosts in the entire sample are classified as star-forming galaxies ($\sim 84 \%$), while $\sim 6\%$ are transitioning and $\sim 10\%$ are quiescent. We draw errors on the fractions for the various sub-samples to determine their statistical significance by randomly drawing ``galaxies" from the full sample (choosing the number of galaxies based on the size of the group), calculate the star-forming fraction, and repeat this process $10,000$ times. We find a slightly higher star-forming fraction for the Gold Sample of $87.8 \%$ (transitioning fraction of $2.4 \%$ and quiescent fraction $9.8 \%$). However, we find a 68\% confidence interval of $\pm 5.7 \%$ from the random draws, thus it is consistent with the full sample. This would imply that the most robust host associations are a fair representation of all included hosts. We find slightly different star-forming fractions compared to the full sample in the Silver Sample ($\sim 73.6 \%$) and more consistent fractions in the Bronze sample ($\sim 89 \%$; Figure~\ref{pie}). To check whether the differences are statistically significant, we draw 19 galaxies (the number of Silver Sample hosts) from the full sample, and find a star-forming fraction of $\approx 84 \pm 8 \%$ (68\% confidence); the Silver star-forming fraction is just below the $68\%$ confidence interval. This could be because Silver Sample hosts are more offset from their host galaxies \citep{BRIGHT-I}, and larger physical offsets are noticed in non star-forming hosts (see Section \ref{sec:offset} for more details). When drawing from nine galaxies (the number of Bronze Sample hosts) from the full sample, we find a median star-forming fraction of $\approx 88 \pm 12 \%$ (68\% confidence), consistent with the full, Gold, and Silver Samples. We find generally that the differences in the distributions are not statistically significant, demonstrating that the distribution in host galaxy type is not strongly dependent on the robustness of host association. However, the difference in the star forming fraction of the Silver Sample may signify that quiescent and transitioning hosts have higher values of $P_\textrm{cc}$'s due to their faintness and/or the short GRB offset. 

We also find that the photometric redshift sample contains a much higher percentage of star-forming galaxies ($\sim 95 \%$) than the full sample and the spectroscopic redshift sample. As there are 20 photometric redshifts, we do find that the star-forming fraction is out of the 68\% confidence interval. As shown in Figure \ref{redshift}, the photometric sample covers a significantly higher redshift range, and the difference in the star-forming fraction is likely due to the prevalence of star-forming galaxies at $z>1$. From the photometry in \citet{BRIGHT-I}, we find that many of these galaxies have little color variation, or difference in apparent magnitudes, between photometric bands, signaling no clear spectral breaks, typically a trait of actively star-forming galaxies. These faint galaxies, if truly at higher redshifts, are thus likely only visible due to the young, massive stars that dominate the SED light in the rest-frame UV \citep{Conroy2013SED}.

\begin{figure*}[t]
\makebox[\textwidth][c]{\includegraphics[width=1.0\textwidth]{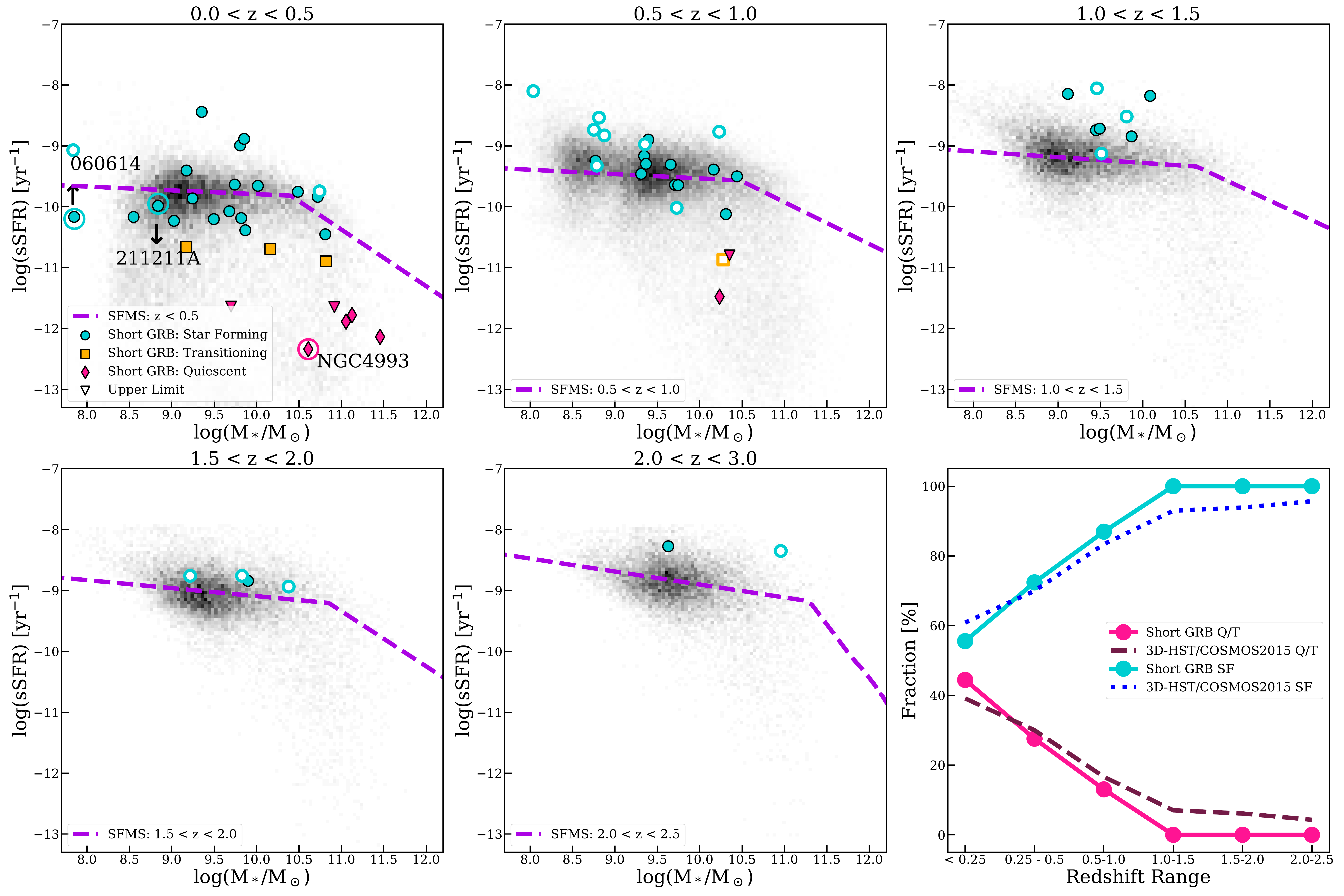}}
\caption{3D-HST and COSMOS2015 \texttt{Prospector} derived sSFRs and stellar masses \citep{Brammer2012, Momcheva2016, Laigle2016, Leja2021} compared to the short GRB host \texttt{Prospector}-derived sSFRs and stellar masses over $0.0 < z < 3.0$. The 3D-HST and COSMOS2015 data is plotted as a normalized histogram in grey. We use the definitions of star-forming (blue circles), transitioning (yellow squares), and quiescent (red triangles) galaxies from \cite{Tachella2021} to plot the short GRB host data. Upper limits are shown as triangles and hosts with photometric redshifts are shown as white-filled shapes. We also circle the host of GRB 170817 (NGC4993). We plot the broken-power law SFMS equations derived in \cite{Leja2021} for each redshift bin. Across all redshifts, short GRB hosts track the SFMS well, implying they are good tracers of star formation given their stellar mass. The bottom right plot shows the star-forming (SF) and quiescent/transitioning (Q/T) fractions of short GRBs in comparison to those of the 3D-HST and COSMOS2015 field galaxies. The short GRB host star-forming and non star-forming fractions are comparable to the field galaxy's fraction at all redshifts, with a slight dominance of non star-forming hosts at $z < 0.25$.}
\label{SFMS}
\end{figure*}

Previous studies based on smaller samples of events used the ``early-type" and ``late-type" classifications, or morphological classifications such as elliptical and disk-dominated. \citet{lb10} found that $\approx 60 \%$ of their 19 short GRB hosts are late-type galaxies and \citet{fbc+13} finds that $\approx 61 \%$ of their 26 hosts are late-type (which includes the 19 burst mentioned in the former work). We find that when we draw 19 and 26 ``galaxies" from our population $10,000$ times the \citet{lb10} and \citet{fbc+13} star-forming fractions are both outside of the $68\%$ confidence interval. The larger star-forming fractions found here are likely due to the much larger host sample, inclusion of a significant number of photometric redshift bursts at $z>1$, and the more quantitative and uniform classification used in this work. Theoretical simulations of BNS mergers, such as that of \citet{csl22}, which also uses the early- and late-type definitions, find similar star-forming host fractions of  $\approx 81-84 \%$ . \citet{mlt+21} predicts a lower star-forming fraction for BNS and NSBH hosts ($\approx 60-70\%$); however they include simulated hosts that are too faint to be detected with ground-based imaging. It is still, however, likely that a majority of the non-host associated GRBs described in \citet{BRIGHT-I} are too faint because they exist at higher redshifts, where star-forming galaxies dominate. Overall, our work is consistent with previous studies that star-forming galaxies make up a majority of short GRB hosts. 

\subsection{Star-Forming Main Sequence}
\label{sec:sfms}
We now determine how well short GRBs trace the normal field galaxy population by comparing their SFRs and stellar masses to the SFMS. The SFMS is a well-studied, redshift-dependent galaxy correlation that is observed to be followed by star-forming galaxies as they gain stellar mass, and is where galaxies spend a majority of their lifetime \citep{Speagle2014, Whitaker2014, Leja2021}. By comparing a host population to the SFMS, we can infer how the short GRB progenitor population traces a combination of stellar mass and star formation in galaxies. In the case of short GRBs, this can inform how galaxy-targeted searches following GW events could be performed (Section \ref{gwfollowup}).

The SFMS is often parameterized by a single or broken power-law in sSFR-$M_*$ space \citep{Speagle2014, Whitaker2014, Leja2021}. Here, we use Equations 9 and 10 and the ``ridge" values listed in Table 1 in \citealt{Leja2021}. This relation provides an excellent comparison to our data set as it is computed from the \texttt{Prospector}-derived SFRs and stellar masses of the COSMOS2015 ($0.2 < z < 0.8$; \citealt{Laigle2016}) and 3D-Hubble Space Telescope (3D-HST; $z > 0.5$; \citealt{Skelton2014}) galaxy surveys. The COSMOS2015 and 3D-HST photometric surveys contain $> 67,000$ galaxies, all with $\geq 17$ bands of photometry, including the NIR wavelengths, which allows for more accurate determinations of these properties. The inferred stellar population properties from the galaxies in these surveys make an ideal comparison set as SFRs and stellar masses, and thus the SFMS, are determined across the range of redshifts relevant for our short GRB sample ($0.2 < z < 3.0$). Instead of using SFR vs. $M_*$ to describe the SFMS, we use sSFR vs. $M_*$, as sSFR normalizes the amount of star formation per unit $M_*$, which is useful when comparing galaxies over a wide range of stellar masses.

We show the comparison of the short GRB host population to the SFMS and COSMOS2015 and 3D-HST field galaxies in Figure \ref{SFMS}, divided into five redshift bins ($z < 0.5$, $0.5 < z < 1.0$, $1.0 < z < 1.5$, $1.5 < z < 2.0$, and $2.0 < z < 3.0$). We find that across all redshifts, short GRB hosts tend to populate the entire SFMS, demonstrating that short GRBs are good tracers of star formation given their host stellar masses. We find that star-forming hosts are in the range of $7.8 \lesssim \log(M_*/M_\odot) \lesssim 10.9$, quiescent hosts are in the range $9.7 \lesssim \log(M_*/M_\odot) \lesssim 11.6$, with transitioning hosts generally between the two populations (Table \ref{tab:allgroup}). We also note that \citet{Leja2021} uses a nonparametric SFH to determine the SFMS. Were we to use the same SFH, the masses would likely increase by $\sim 0.2$~dex and sSFR would decrease slightly, with more significant differences at higher redshifts \citep{Leja2019}. However, these differences would not be significant enough to change the trend noticed between short GRB hosts and the SFMS; however, might explain why short GRB hosts at higher redshifts tend to lie above the SFMS.

For comparison, we also include NGC4993, the quiescent host of GW170817/GRB\,170817A \citep{gw170817, bbf+17, llt+2017, pht+17, Pan2017}. While NGC4993 is not the most massive host compared to the short GRB host sample at $z<0.5$, it is by far the most quiescent. We find that compared to the short GRB host sample, GW170817 occurred in a unique environment, as the host has low star formation (sSFR $\approx 10^{-12}$ yr$^{-1}$; $\approx 10^{-3}$yr$^{-1}$ less than the entire host population and $\approx 10^{-1}$yr$^{-1}$ less than the non star-forming hosts) for its stellar mass ($M_* \approx 4.1\times10^{10} M_\odot$). The majority of galaxies at this stellar mass and redshift ($\approx 75 \%$ in the 3D-HST sample) are expected to be star-forming. We also compare GRB the hosts of GRBs 060614 and 211211A to the host population, as they are low-redshift long GRBs with significant evidence for a NS merger origin  \citep{dcp+2006, gfp+06, jgl+2022}. Both hosts have some of the lowest masses and SFRs compared to the short GRB host population (bottom $12\%$ and $19\%$, respectively). However, they are consistent with the short GRB host population. Their stark contrast to the stellar mass, sSFR, and optical luminosity \citep{BRIGHT-I} of NGC4993 highlights the range of environments of compact object mergers. 

In the final panel of Figure \ref{SFMS}, we show the fraction of quiescent and transitioning galaxies compared to the 3D-HST and COSMOS2015 datasets. We use the stellar mass and sSFR data in \citet{Leja2019} to determine the quiescent and transitioning fractions of the field galaxy population with the same method described in Section \ref{sec:sf_class}. We find that the short GRB host fractions are comparable with the field galaxy population at all redshifts, and begin to dominate the population at $z < 0.25$ (although numbers are small in this regime). As galaxies at $z < 0.25$ are easily observable, we suspect that this does represent the true rate of short GRBs in non star-forming galaxies. Since quiescent hosts also have larger mass-weighted ages, this population could be indicative of a higher fractions of long delay time progenitors. The lack of quiescent and transitioning galaxies at high redshifts ($z > 1.0$) is likely due to observational limitations, as these galaxies are much fainter and thus much more difficult to detect with ground-based imaging.

Overall, we find short GRB hosts tend to have the expected stellar mass given their SFRs and thus tend to trace the SFMS. Furthermore, the fractions and stellar masses for star-forming, transitioning, and quiescent hosts also are within the observed range of field galaxies at their respective redshifts.

\begin{figure*}[t]
\vspace{-2in}
\makebox[\textwidth][c]{\includegraphics[width=1.2\textwidth]{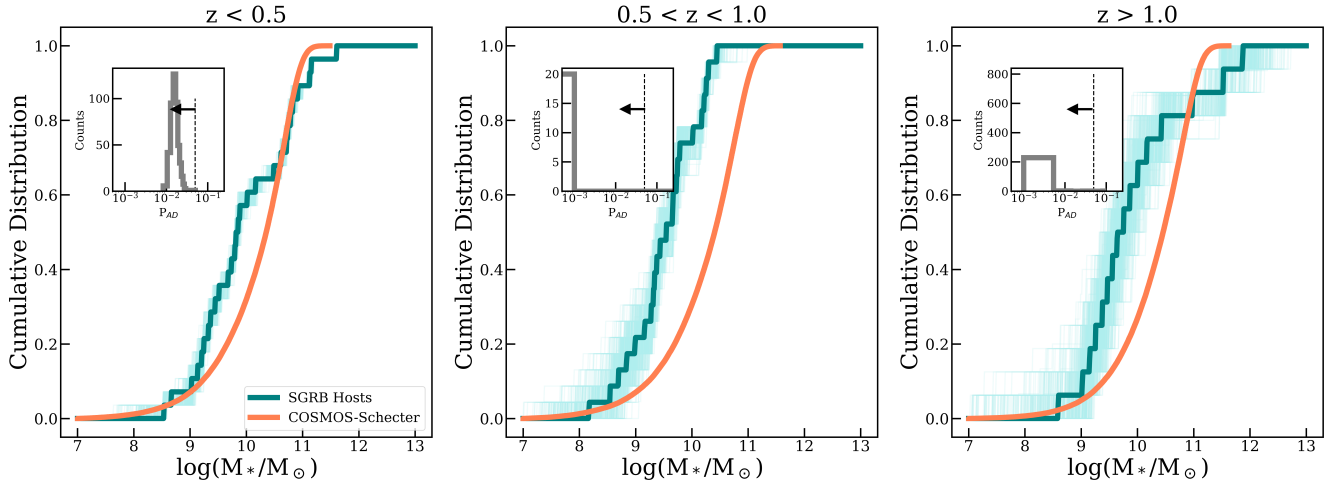}}
\vspace{-2.1in}
\caption{CDFs of the short GRB host stellar masses (teal) compared to the mass-weighted Schechter function of the COSMOS survey's active galaxies (orange; \citealt{Davidson2017}) across three redshift bins spanning $0 < z < 3.0$. For the short GRB host mass distributions, we plot 5000 realizations on the CDF along with the median CDF (teal solid line). The Schechter function is used to represent the luminosity-corrected distribution of normal field galaxy stellar masses. {\bf Insets:} We determine if the short GRB host and the mass-weighted Schechter functions are drawn from the same underlying distribution through an Anderson-Darling (AD) test; the distribution of $P_{AD}$ values for all realizations. The dashed line in the insets shows $P_{AD} = 0.05$, with arrows in the direction of the point below which we reject the null hypothesis that they are drawn from the same underlying mass distribution. Across all redshift bins, we can reject the null hypothesis that short GRB hosts trace the stellar mass distribution in galaxies.}
\label{Schechter}
\end{figure*}

\subsection{Stellar Mass Distribution}
\label{sec:mass}
We next compare the short GRB host stellar masses to that of the normal field galaxy population to understand how they trace the stellar mass distribution in galaxies. We divide the host population into three redshift bins ($z < 0.5$, $0.5 < z < 1.0$, and $z > 1.0$) and compute the stellar mass cumulative distribution from 5000 realizations of the \texttt{Prospector}-derived posterior distributions and the medians of those distributions. We use a mass-weighted Schechter function of the COSMOS2015 galaxy survey, detailed in \citealt{Davidson2017}, to represent the stellar mass distribution of field galaxies within the three redshift bins. The Schechter function is an empirical parameterization for the number of galaxies at a given stellar mass, and the \citet{Davidson2017} function ensures that the field galaxy stellar mass distribution is not biased by any observational limits of the COSMOS2015 survey (e.g. correcting for missing low-luminosity or high redshift galaxies). The Schechter function in \citet{Davidson2017} was built from stellar masses produced by a delayed-$\tau$ SFH from the SED modeling code \texttt{GALAXEV} \citep{swc+2014}. As choice of SFH has a greater effect on stellar population property values than other galaxy modeling assumptions, our \texttt{Prospector}-produced stellar masses are comparable to theirs due to common SFH. To test whether the occurrence rate is proportional to stellar mass, we convert the Schechter function into a total mass budget by multiplying it by a uniform distribution of stellar masses between $10^7$ and $10^{13}$\,M$_\odot$ and calculate the CDF. We show the distributions in Figure \ref{Schechter}. We find that the medians of the field galaxy stellar masses in our three redshift bins are higher by $\approx 0.3-0.5$~dex than those of the host distributions.

For a quantitative comparison to the short GRB host sample, we perform AD tests with the null hypothesis that both populations are derived from the same underlying stellar mass distribution. We perform an AD test for each of the 5000 realizations of the CDF and the Schechter function to build a distribution of probabilities ($P_{AD}$), shown in the inset of Figure \ref{Schechter}. For all redshift bins, all $P_{AD} < 0.05$, and, therefore, we can reject the null hypothesis that short GRB hosts trace the stellar mass distribution of field galaxies. Given that there are more high-mass galaxies in the normal field galaxy population than in the short GRB host sample, this also implies that high mass galaxies are not necessarily more likely to host short GRB progenitors. Furthermore, given that short GRB hosts are tracing the SFMS, this indicates that short GRBs do not trace stellar mass alone; instead they trace a combination of star formation and stellar mass.

We note that if the true short GRB host population traces the stellar mass distribution in galaxies, and the discrepancy is purely an observational selection effect, this would require missing a fraction of high-mass hosts. As an exercise, we add a number of additional $\log(M_*/M_\odot) = 10.5$ galaxies to the short GRB sample to determine how many more high-mass hosts are needed for the observed host population to trace the stellar mass distribution. For $z < 0.5$, we find that we would need to nearly double the number of high-mass galaxies (4-6 more $10^{10.5}$ M$_\odot$; $> 50\%$ of $P_{AD} > 0.05$). However, since high-mass galaxies at $z < 0.5$ are easily detectable given their higher luminosities, it is very unlikely that we are missing this many hosts in our sample. If anything, there is an observational bias against identifying low-mass galaxies as short GRB hosts, which would further drive the distributions apart and strengthen our conclusions.

\subsection{Stellar Population Age}
\label{sec:age}

\begin{figure}
\centering
\includegraphics[width=0.5\textwidth]{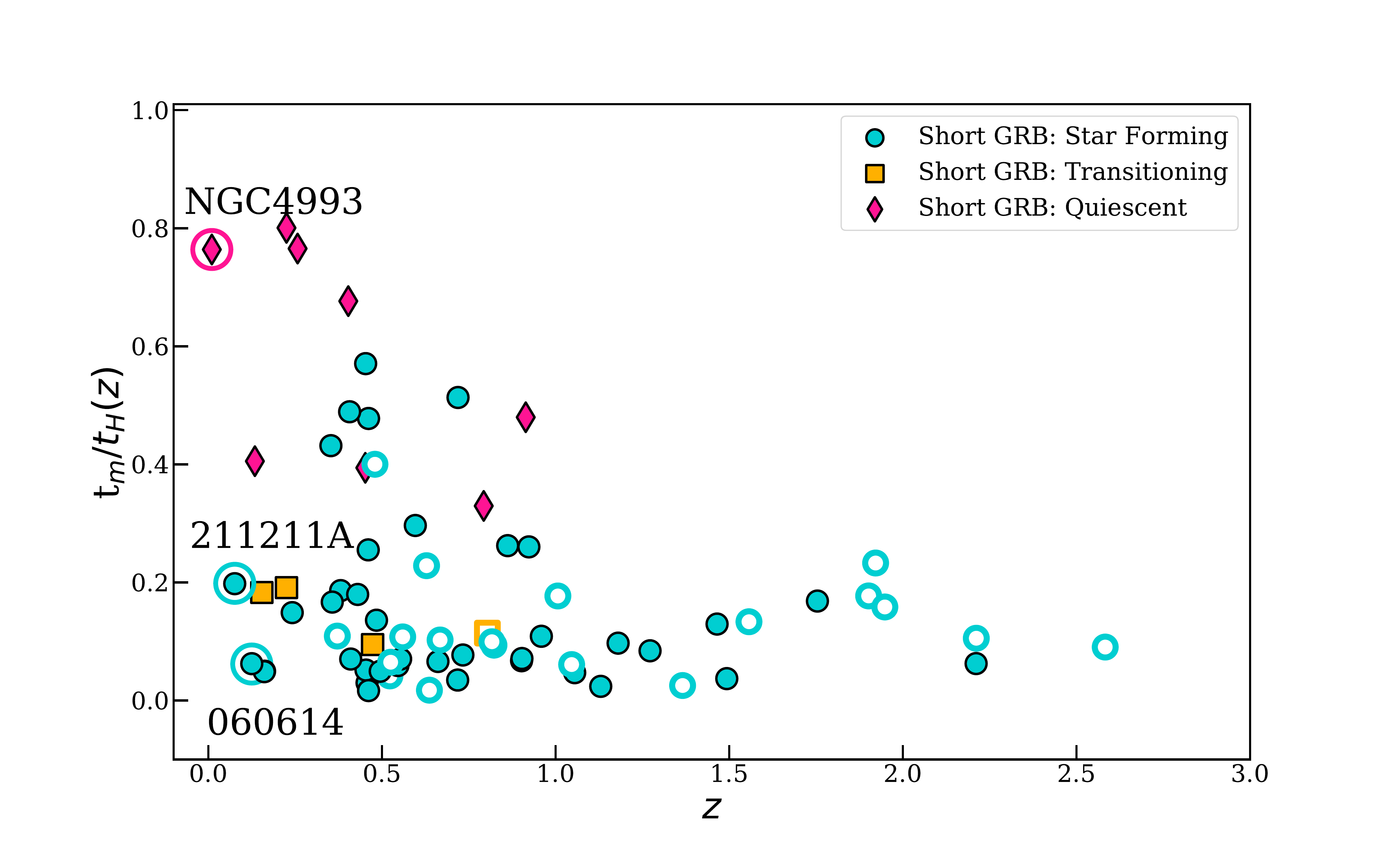}
\vspace{-0.3in}
\caption{The mass-weighted ages ($t_m$) of short GRB hosts normalized by the age of the universe at their redshift ($t_\textrm{H}(z)$) versus redshifts, color-coded by their host type (blue is star-forming, yellow is transitioning, and red is quiescent). White-filled symbols are photometric redshifts and single color symbols are spectroscopic redshifts. We find that closer to the epoch of star-formation ($z=2$), there are more hosts that are much younger than $t_\textrm{H}(z)$. As redshift decreases, we see that the population of hosts begins to be dominated by those with ages closer to $t_\textrm{H}(z)$. This would imply that production of the short GRB progenitors is decreasing with redshift, as the short GRBs found at lower redshifts likely formed closer to the epoch of star formation and have a longer delay time.} 
\label{ageredshift}
\end{figure}
The stellar population age is an important property in understanding short GRB progenitors, as it can be used as a loose proxy of the formation and merger timescales, or the delay time, of the progenitor (e.g., \citealt{sbl+2019}). Indeed, past work has shown that the wide range of stellar population ages is consistent with the range of compact object merger timescales (e.g., \citealt{sb2007, zzr2007, obk08, lb10, ber14}). In Figure~\ref{cdf_pdf}, we show the mass-weighted age ($t_m$) median CDF and 5000 realizations of the \texttt{Prospector}-derived posteriors for all, star-forming, and transitioning and quiescent hosts. As expected, we find a large difference in ages between the star-forming and non star-forming populations; star-forming hosts have $\langle t_m \rangle = 0.63^{+1.86}_{-0.38}$~Gyr (median and 68\% credible interval), whereas non star-forming hosts have $\langle t_m \rangle = 2.87^{+5.22}_{-1.96}$~Gyr. As there are more star-forming hosts in the sample, the entire population is weighted more heavily by the younger hosts and thus has a median age as a population of $\langle t_m \rangle = 0.80^{+2.71}_{-0.53}$~Gyr. We find that the shape of the posterior distributions of the entire host population roughly follows that of a power-law decline, which is the expected shape of the delay-time distribution of BNS mergers \citep{peters1964, Nakar2006,Jeong2010,Hao2013}. Using the age distribution as a proxy for progenitor delay times, this is consistent with previous findings that there exists a broad range of delay times for the expected progenitors of short GRBs, including BNS and NSBH mergers \citep{bkv2002, GuettaPiran2005, Nakar2006, Hao2013, Wanderman2015, Anand2018}. 

\begin{figure}
\centering
\includegraphics[width=0.5\textwidth]{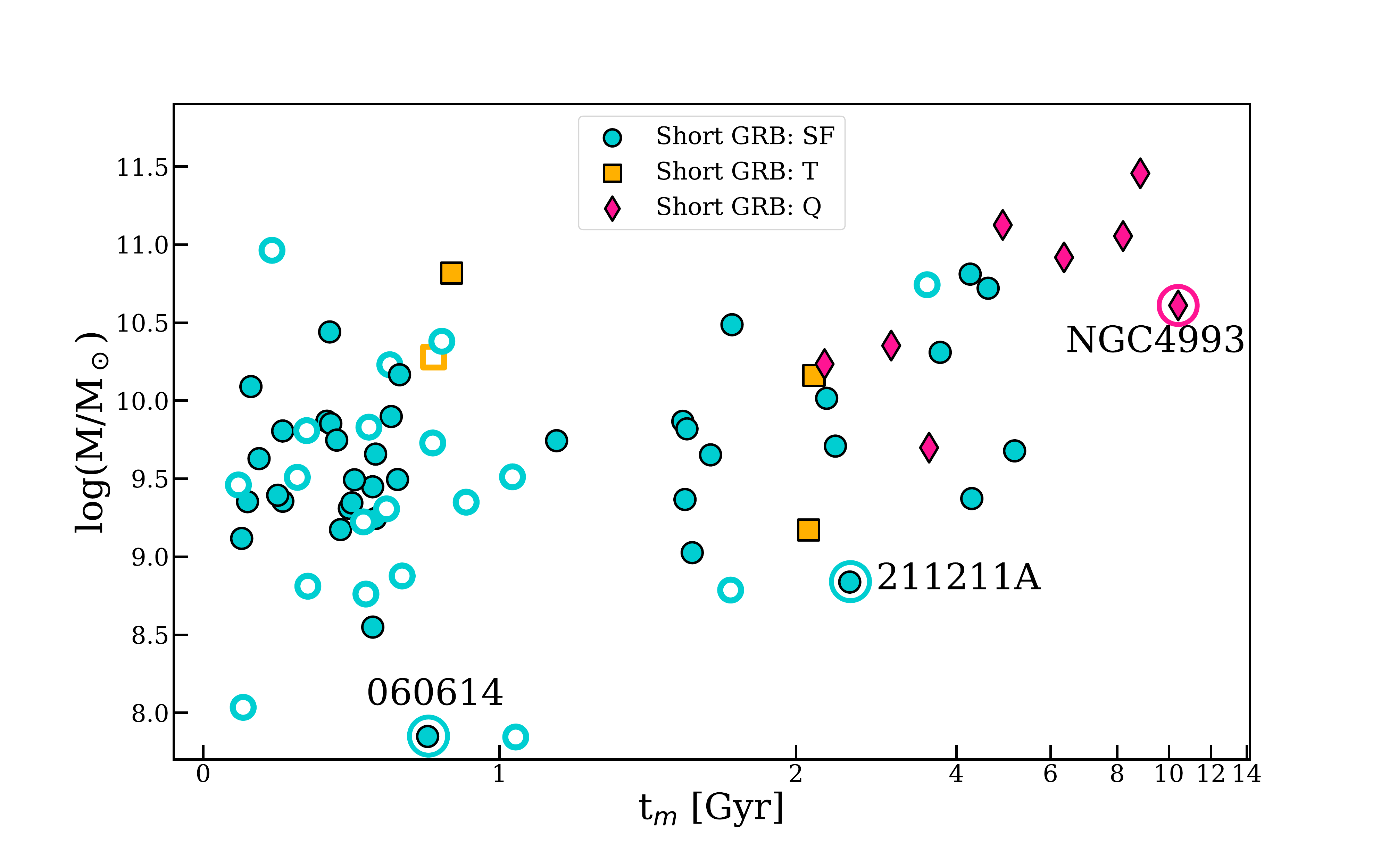}
\vspace{-0.3in}
\caption{Stellar masses and mass-weighted ages of short GRB hosts, with the same color and shape schematic as in Figure \ref{ageredshift}. Hosts with photometric redshifts are represented with white-filled shapes. We mark NGC4993 and the hosts of GRBs 060614 and 211211A  for reference. We find that quiescent host galaxies make up the oldest and most massive hosts in the sample, and that the hosts of short GRBs span nearly the full range of galaxy properties.}
\label{massage}
\end{figure}

To understand age evolution with redshift, we show the stellar population age normalized by age of the universe at each redshift ($t_\textrm{\rm H}(z)$) versus redshift in Figure \ref{ageredshift}. We find that at $z > 1$, the population of hosts is dominated by stellar populations that are only $\approx 20\%$ the age of the universe. We also find that all but three hosts with photometric redshift determinations have ages $< 2$~Gyr, are heavily clustered at $\lesssim 1$~Gyr (Figure \ref{massage}), and are more skewed toward higher redshifts. This is consistent with the association of short GRB progenitors at $z>1$ to more recent star formation, as the youngest stellar population ages are just slightly above the stellar evolutionary timescale of NS progenitors $\mathcal{O}(10$~Myr). At face value, higher redshift host galaxies are closer in time to the epoch of peak star formation of the universe \citep{md14}, and thus the dominance of younger stellar population ages at $z>1$ is likely indicative of the fast-merging binary population comprising of mostly BNS progenitors with tight orbital separations at formation and short delay times \citep{bkv2002, bpp2019, sbl+2019}. This also implies a larger population of fast-merging, $z>1$ BNS systems than previously confirmed (e.g., \citealt{skm+18, pfn+20, otd+2022}), as we have nearly 15 hosts at $z>1.0$. Taking into account completeness, we believe $25-44 \%$ of \textit{Swift} short GRBs are from $z > 1$ \citep{BRIGHT-I}. In contrast, at $z<1$, the ages span a wide range compared to $t_{\rm H}(z)$. In particular, the growing number of older stellar populations with $t_m/t_{\rm H}(z) \approx 0.8$ indicates that these represent hosts of binary systems with longer delay times that formed during the epoch of peak star formation but did not merge until much more recently (Figure~\ref{ageredshift}). We note that the longer delay times are most probably derived from the intrinsic properties of isolated binary systems, including orbital separation, rather than delayed mergers from dynamical formation in e.g. globular clusters, as it is highly unlikely for these BNS systems to merge within a Hubble time \citep{bkd02,yfk+2020}. We also find a much wider range of $t_m/t_H$ values at $z<1$ than at $z>1$, consistent with the fact that binary production has decreased and we are no longer dominated by very fast merging channels at $z<1$. Taken together, we find this likely implies that BNS/NSBH production has decreased since $z\approx1$, as the population in the near universe is no longer dominated by fast-merging systems as at higher redshifts. We use these properties to quantify the DTD parameters in \citet{Zevin+DTD}. A similar DTD is observed with the Galactic BNS population in which there are many more binaries expected to merge quickly ($< 1$~Gyr; \citealt{bpp2019}).

We also explore the ages in the context of host stellar mass, which we show in Figure \ref{massage}. As expected, we find that as the host age increases, the host mass also increases and that quiescent hosts comprise the oldest and most massive galaxies in the sample. This is in alignment with standard galaxy evolution, as galaxies gain in mass as they evolve, through major and minor mergers \citep{cmw2011, Whitaker2014}. It is unclear if galaxy mergers have an effect on binary production, although NGC4993 (GRB 170817A) did have a recent galaxy merger \citep{pht+17, eby+2020, kfb+2022}. Thus, the wide range of short GRB host masses distribution reinforces the wide distribution of delay times possible for their progenitors. Overall, short GRBs originate in galaxies spanning the full range of properties. Although the parametric SFH assumption is known to produce younger ages than nonparametric SFH models by a factor of $\approx 3-5$ \citep{pdf+2001, clj+2019}, this will not affect the relative fraction of young and old stellar populations.   

Overall, we find that the mass-weighted ages of short GRB hosts reinforce a wide range of progenitor delay-times, with more support for fast-merging systems at $z > 1$, and a decrease in binary production toward lower redshifts (see \citealt{Zevin+DTD} for more). Ground-based gravitational wave detectors will not be able to reach these cosmological distances for $\gtrsim 15$ years. Therefore, they will not be able to fully sample the DTD of BNS/NSBH mergers, such as is possible with short GRBs.

\begin{figure}[t]
\centering
\includegraphics[width=0.49\textwidth]{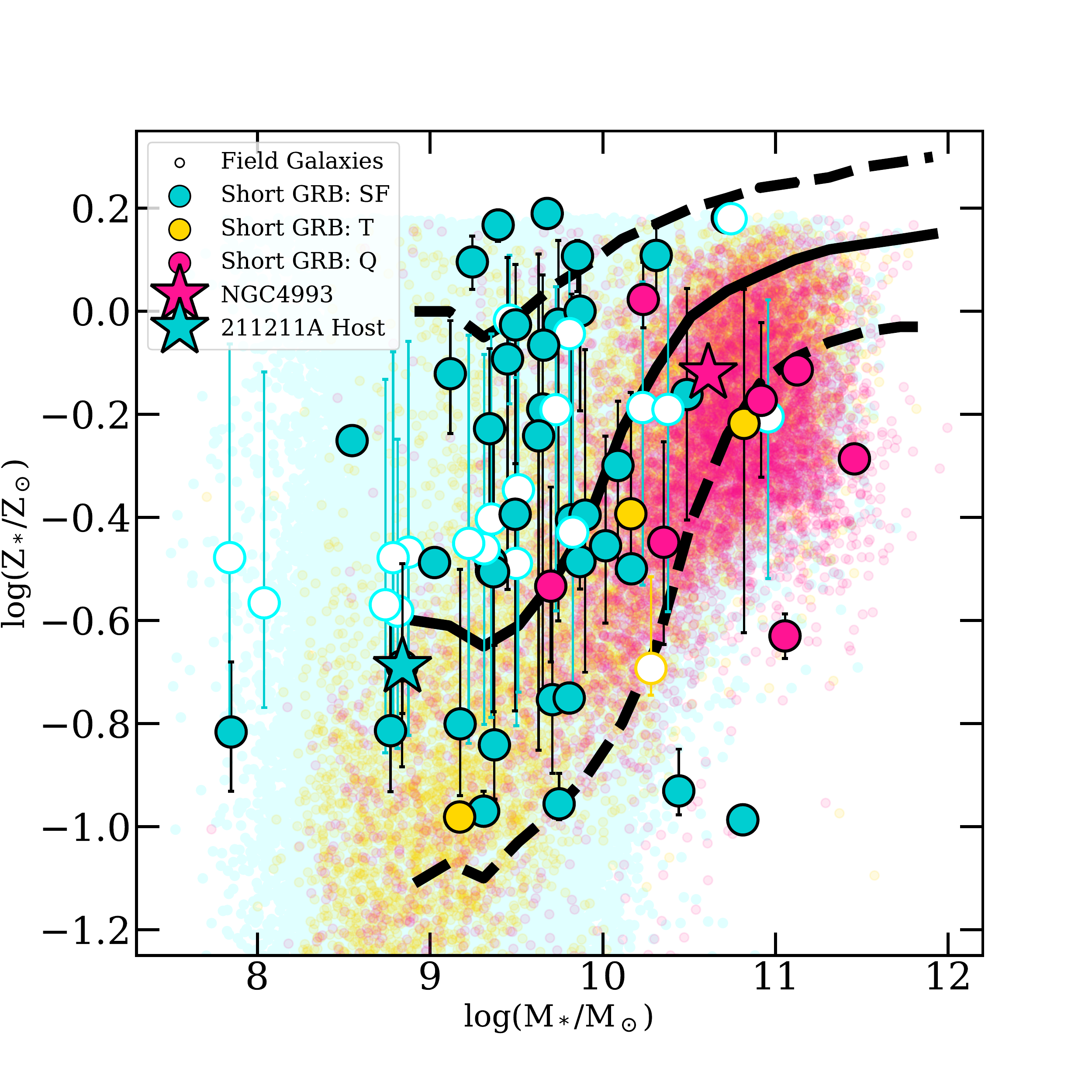}
\vspace{-0.35in}
\caption{The \citet{gcb+05} stellar mass-stellar metallicity relation (MZ; solid black lines represent the median and dashed black lines represent the 68\% credible region) overplotted with short GRB hosts. The colors of both the 3D-HST and COSMOS field galaxies (small circles) and the short GRB hosts (large, outlined circles) denote their star formation classification, with blue representing star-forming, yellow representing transitioning, and pink representing quiescent galaxies. The open circles represent photometric redshift hosts. Short GRB hosts follow the MZ relation, fall within the expected range for their star formation classification, and span a wide range of metallicities. We also highlight NGC4993 and the host of GRB 211211A for reference, which occur in very different places in the MZ-plane. Overall, GRB-producing NS mergers form independently of the metallicity of their stellar populations. }
\label{MZrelation}
\end{figure}

\subsection{Mass and Stellar Metallicity}
\label{sec:metallicity}
Finally, we explore the stellar mass-stellar metallicity relation of short GRB hosts. Previous studies on the effects of binary compact object merger formation and metallicity have found that lower metallicity environments during star formation lead to more compact binaries \citep{bkd+2010, obb+2017}. Stellar metallicity has been shown to trace both the star formation and age of galaxies, with more evolved stellar populations being more metal-rich, thus giving insight into the types of stars in an environment \citep{gcb+05, Panter2008}. Studies have also shown that stellar metallicity increases with increasing stellar mass, as more massive galaxies contain older stellar populations and thus more metals. 

In Figure \ref{MZrelation}, we show the derived log(M$_*$/M$_\odot$) and log(Z$_*$/Z$_\odot$) in comparison to the \citet{gcb+05} mass-metallicity (MZ) relation, as well as the COSMOS2015 and 3D-HST field galaxies color-coded by their star formation classification. We find that most short GRB hosts fall within the expected metallicity range given their stellar mass and star formation classification. Figure~\ref{MZrelation} also shows that BNS/NSBH mergers are able to form over a wide range of metallicity environments, even in the local universe; this is highlighted with the low-redshift hosts of GRBs 170817A (NGC4993) and 211211A. This suggests that short GRB formation efficiency across all redshifts is not dependent on the amount of metals in their environment. If short GRBs only occurred in low-metallicity environments, similar to long GRBs and SLSNe, we would witness the entire population very connected to recent star formation. Conversely, a dependence on higher metallicity hosts would suggest that short GRBs overwhelmingly come from less star-forming populations. The spread in host metallicities therefore shows that short GRBs derive from both fast-merging populations linked to recent star-formations and slowly-merging progenitors not linked to star-formation. Furthermore, it shows that the short GRB progenitor formation efficiency is also independent of metallicity which is consistent with predictions for BNS systems \citep{cmb+2018, gm2018, nvs+2019}. However, NSBH systems have a higher formation efficiency in lower metallicity environments \citep{kmg+2018, bbs+2021}; thus, the MZ relation of short GRB hosts does not follow expectations if all events are derived from NSBH mergers.  

\section{GRB and Locations versus Host Properties}
\label{sec:grb_prop}
We next explore how GRB properties, such as $T_{90}$, prompt fluence (15-150 kev), presence of detectable extended emission, optical afterglow detection, afterglow luminosities, and galactocentric offsets compare with their host properties to understand if host properties influence the GRB properties.

\subsection{Duration, Fluence, and Extended Emission}

\begin{figure}
\centering
\includegraphics[width=0.475\textwidth]{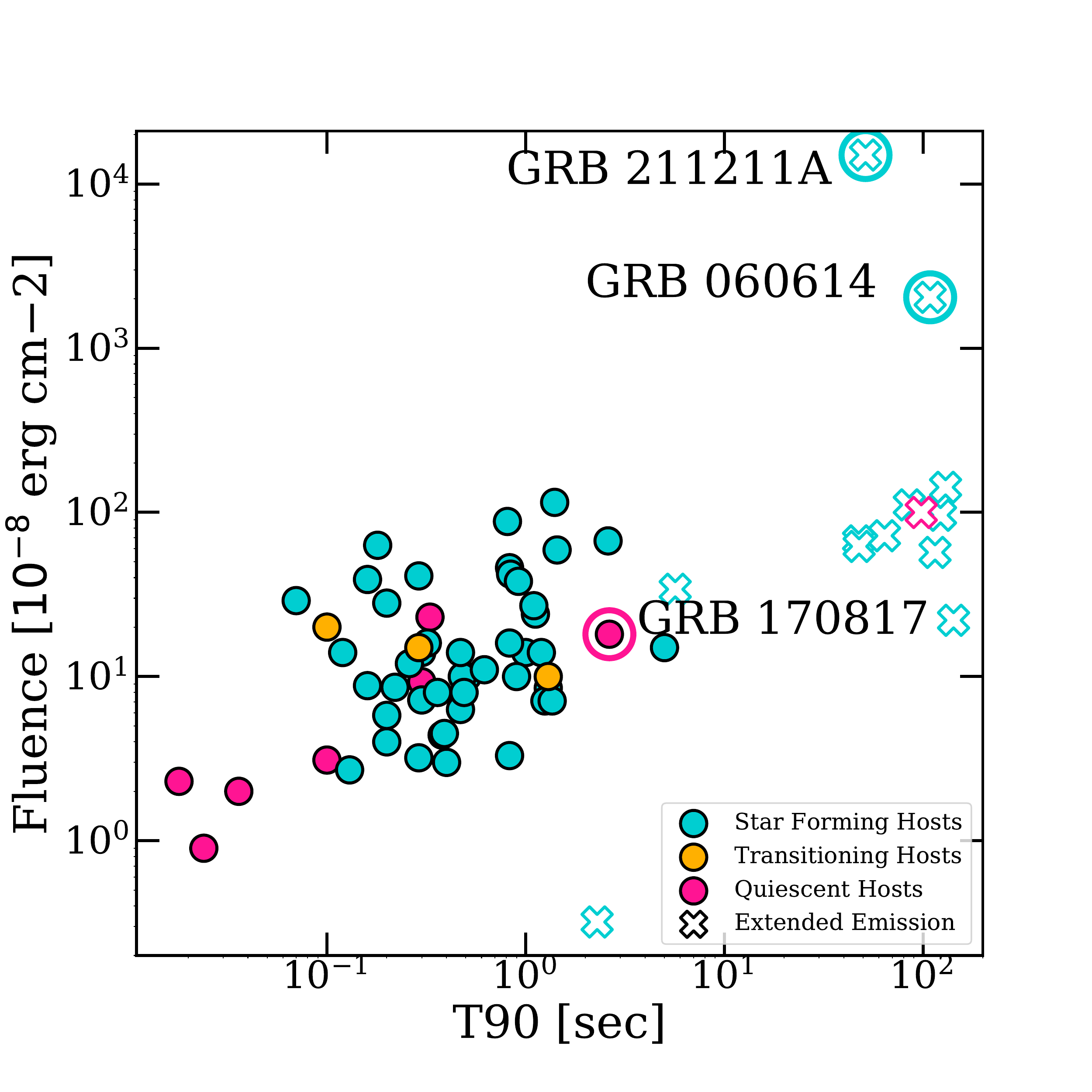}
\vspace{-0.1in}
\caption{The $T_{90}$ and fluences of short GRBs, marked by the star formation classification of their hosts. We see that all types of hosts (star-forming, transitioning, and quiescent) span the full ranges, implying that host type does not influence these GRB properties. Short GRBs in quiescent hosts, however, do tend to have the lowest fluences. We mark GRBs 060614, 170817, and 211211A for reference.}
\label{t90fluence}
\end{figure}

We first compare host type (star formation classification) to $T_{90}$ and fluence in Figure~\ref{t90fluence}. We find values for $T_{90}$ and fluences in the {\it Swift} BAT3 catalog\footnote{\url{https://swift.gsfc.nasa.gov/results/batgrbcat/summary_cflux/summary_general_info/summary_general.txt}} \citep{lsb+16}. While short GRBs are typically defined with $T_{90} \lesssim 2$~s, nominally at the $T_{90} \approx 2$~s boundary between short and long GRBs (which derive from young, massive stellar progenitors and are known to only occur in star-forming galaxies; \citealt{lb10, Svensson2010, Perley2013, Wang2014, Vergani2015, Niino2017}), we might expect some contamination in our sample. In Figure \ref{t90fluence}, we find that for very short durations, $T_{90}\lesssim 0.1$~sec, and low fluences, $\leq 3.1\times 10^{-8}$~erg cm$^{-1}$, almost all hosts are quiescent, while almost none are star-forming in this region. Since $\gamma$-ray fluence has been shown to correlate with X-ray and optical afterglow luminosity \citep{gbb+08}, this could be a product of the lower ISM densities of quiescent hosts. However, including GRB\,170817A and the extended emission short GRB\,050724A, quiescent hosts span the full range of parameter space. The interval of bursts with $T_{90}\approx 0.1-2$~s is fully occupied by star-forming and transitioning hosts, except for the quiescent hosts of GRBs 100625A and 100117, (Figure \ref{t90fluence}) but span two orders of magnitude in fluence. Thus, while we do not find any compelling evidence for significant contamination from massive star-originated events (see also \citealt{BRIGHT-I}), we do note the prevalence of quiescent hosts with lower fluences and shorter durations.

\begin{figure*}[t]
\makebox[\textwidth][c]{\includegraphics[width=1.0\textwidth]{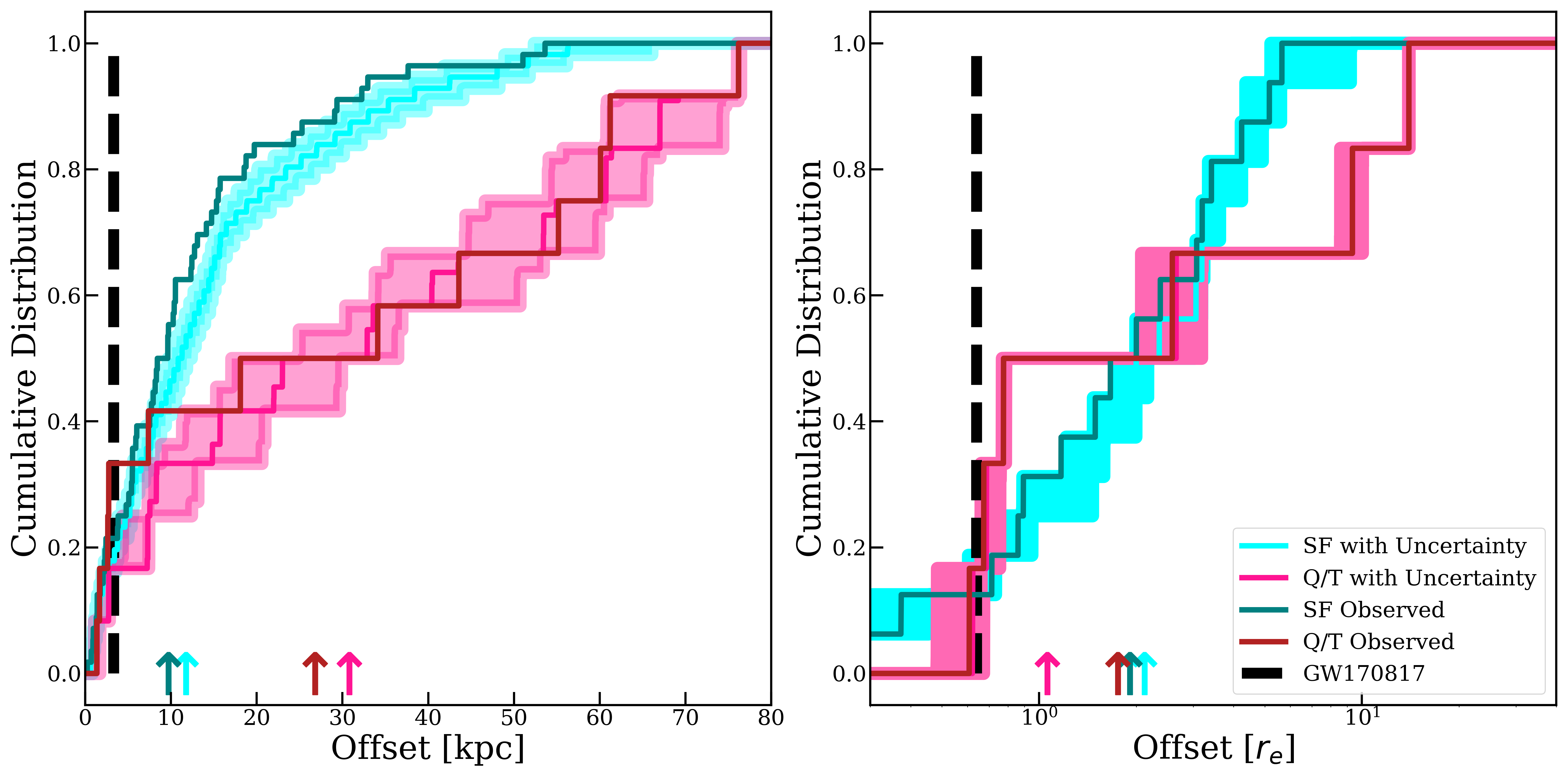}}
\vspace{-0.2in}
\caption{CDFs of observed physical (\textit{left}) and host-normalized (\textit{right}) offsets for star-forming (dark blue) and quiescent/transitioning (red) hosts. The CDFs produced from the Rice distributions are shown in the respective lighter colors for each host star formation classification. We show medians with the arrows and highlight the offset of GRB\,170817 with a dashed black line. We find significant differences in physical offsets between the star-forming and non star-forming hosts, but more similar distributions when looking at the host-normalized offsets. Taken together, this likely implies that short GRB progenitors in non star-forming galaxies have migrated further than those in star-forming hosts.}
\label{offsets}
\end{figure*}

We further investigate how the hosts of extended emission bursts compare to those of the normal short GRB host population. Several events in our sample have extended emission detectable in the $\gamma$-rays which results in $T_{90} \gg 2$~s \citep{nb06,lsb+16,BRIGHT-I}\footnote{These GRB are 050724, 061006, 060614, 061210,  070714B, 080123, 090510, 170728B, 180418A, 180618A, 180805B, and 200219A.}. While the physical origin of extended emission is uncertain, it has been suggested that they may be derived from NSBH merger systems, on the basis of their potentially smaller galactocentric offsets, and larger $\gamma$-ray energies \citep{tko+08, glt2020}. By default, extended emission bursts share similar $\gamma$-ray properties to long GRBs given their long durations and larger fluences, but unlike long GRBs, they are not accompanied by Type Ic SNe. We find that the normal short GRB host population and distribution of extended emission bursts are very similar in stellar mass and age. For the stellar masses and mass-weighted ages, we find $p_{\rm AD} > 0.25$ between the two populations, and thus cannot negate that they are derived from the same underlying mass and age distributions. We also do not find any difference between the star formation classifications between each population: we randomly draw 12 hosts (the number of extended emission hosts) from the normal short GRB host population and find a star-forming fraction of $83.3 \% \pm 10.5$ ($1\sigma$). This is consistent within the $68\%$ credible interval to the observed star-forming fraction of extended emission hosts of 91.6\%. Thus, consistent with the conclusions of \citet{BRIGHT-I}, the host properties alone do not clearly support a distinct progenitor for extended emission bursts.

\subsection{Galactocentric Offsets}
\label{sec:offset}

We next explore the short GRB offsets compared to star formation classification, host stellar mass, and stellar age. \citet{nfd+20} found that the short GRBs in galaxy clusters (GRBs 050509B, 090515, and 161104A), which host the oldest and most massive galaxies in the universe, were significantly more offset from their hosts than the rest of the population, suggesting that the host type or age might affect the progenitor's offset (see also: \citealt{bsm+07, sb2007}). We compare the projected physical (in kpc) and host-normalized offsets  (in effective radii units $r_e$) from \citet{BRIGHT-I} of the star-forming and non star-forming (quiescent and transitioning) hosted short GRBs in Figure \ref{offsets}. We examine both the observed offset CDF and offset distribution including uncertainties, which is built from 5000 realizations of a Rice distribution (Equation 2 in \citealt{bbf16}; see Section~6.2 of \citealt{BRIGHT-I}) using the offsets and their $1\sigma$ uncertainties in \citet{BRIGHT-I}. We find that the short GRBs in star-forming hosts have median offsets of $9.01^{+11.62}_{-6.82}$~kpc and $1.83^{+2.08}_{-1.06} r_e$ (sampled distributions give $11.04^{+17.31}_{-8.61}$~kpc and $2.04^{+2.2}_{-1.34} r_e$). Meanwhile, short GRBs in non star-forming hosts have larger observed offsets of $26.1^{+34.26}_{-23.70}$~kpc and $1.68^{+8.59}_{-1.02} r_e$ (sampled distributions give $30.09^{+35.57}_{-27.27}$~kpc and $1.02^{+12.91}_{-0.35} r_e$); Figure~\ref{offsets}. 

Overall, the short GRBs in non star-forming hosts are observed to be farther from their hosts than short GRBs in star-forming hosts (Figure~\ref{offsets}; $p_{\rm AD}=0.004$). However, we find that the differences are less clear for host-normalized offset distributions ($p_{\rm AD}=0.25$). The observed offsets result from a combination of the systemic velocity of the binary progenitor and its delay time \citep{fwh99}. Given similar systemic velocities of binary progenitors in both star-forming and non star-forming hosts, the observed physical offset differences can be naturally explained by the longer delay times in non star-forming hosts (commensurate with their older ages; c.f., Figure~\ref{cdf_pdf}). Galaxies can also grow significantly over approximately gigayear timescales, and thus short GRBs from non star-forming hosts may have gotten kicked when the galactic potential was much different than it is today. However, even if all short GRB progenitors migrated an approximately equal physical distance from their host centers, this would translate to less distance traversed within the quiescent or transitioning hosts, as their radii are larger. Rather, the differences in host-normalized offsets between star-forming and non star-forming hosts are not expected to be as prominent as in physical space. Finally, we note that we only have host-normalized offsets for the most secure host associations, thus creating a slight selection effect toward bursts at smaller offsets.

\subsection{Afterglow Properties}
We also examine if host properties are correlated with short GRB optical afterglow detectability and luminosities. Naively, we expect that star-forming hosts will have higher ISM densities on average, and, thus, brighter optical afterglows (cf., \citealt{gs02}) than those in quiescent or transitioning hosts. However, we see no difference in distributions between optically detected and non-detected GRBs, finding $P_{AD} > 0.24$ for all tested properties. 

We further compare the optical afterglows of short GRBs to their host types. We collect optical afterglow luminosities or upper limits from several minutes to several months after the burst (\citealt{fbm+15,rfk+2021} and references therein), and convert to luminosities in the host frame using the redshifts found in this work. Here, optical refers to afterglow detections taken in optical and near-infrared photometric filters, which de-redshifted lie in the optical wavelengths, approximately at the same rest-frame wavelength. In total, we have 49 optical afterglow limits and detections, 82.4\% coming from short GRBs in star-forming galaxies, 7.8\% in transitioning, and  9.8\% in quiescent, which roughly matches the breakdown of our full sample. We compare all afterglows at a common rest-frame time of $t_{RF}=5$~hr, as the afterglow has not significantly faded, interpolating or extrapolating from their nearest detection or deepest upper limit when necessary. We build Gaussian distributions of the flux of the detection using their median and $3\sigma$ uncertainties. For the $3\sigma$ upper limits, we build a top-hat function from a lower limit of $10^{38}$ erg/s (representing $\gtrsim 100$ times lower than the faintest detected afterglow) to the value of the upper limit. We draw randomly from these distributions 1000 times and build CDFs of the luminosities based on whether the galaxy is star-forming or not. We show the median and 1$\sigma$ confidence interval of these distribution in Figure \ref{AG}. While it appears that short GRBs in quiescent and transitioning hosts have fainter afterglow luminosities than those in star-forming hosts, we cannot rule out the null hypothesis that they are drawn from the same underlying distribution ($p_{\rm AD} = 0.24^{+0.01}_{-0.16}$). Thus, we do not find statistical support for fainter optical afterglows in non star-forming host galaxies.

\begin{figure}
\centering
\includegraphics[width=0.45\textwidth]{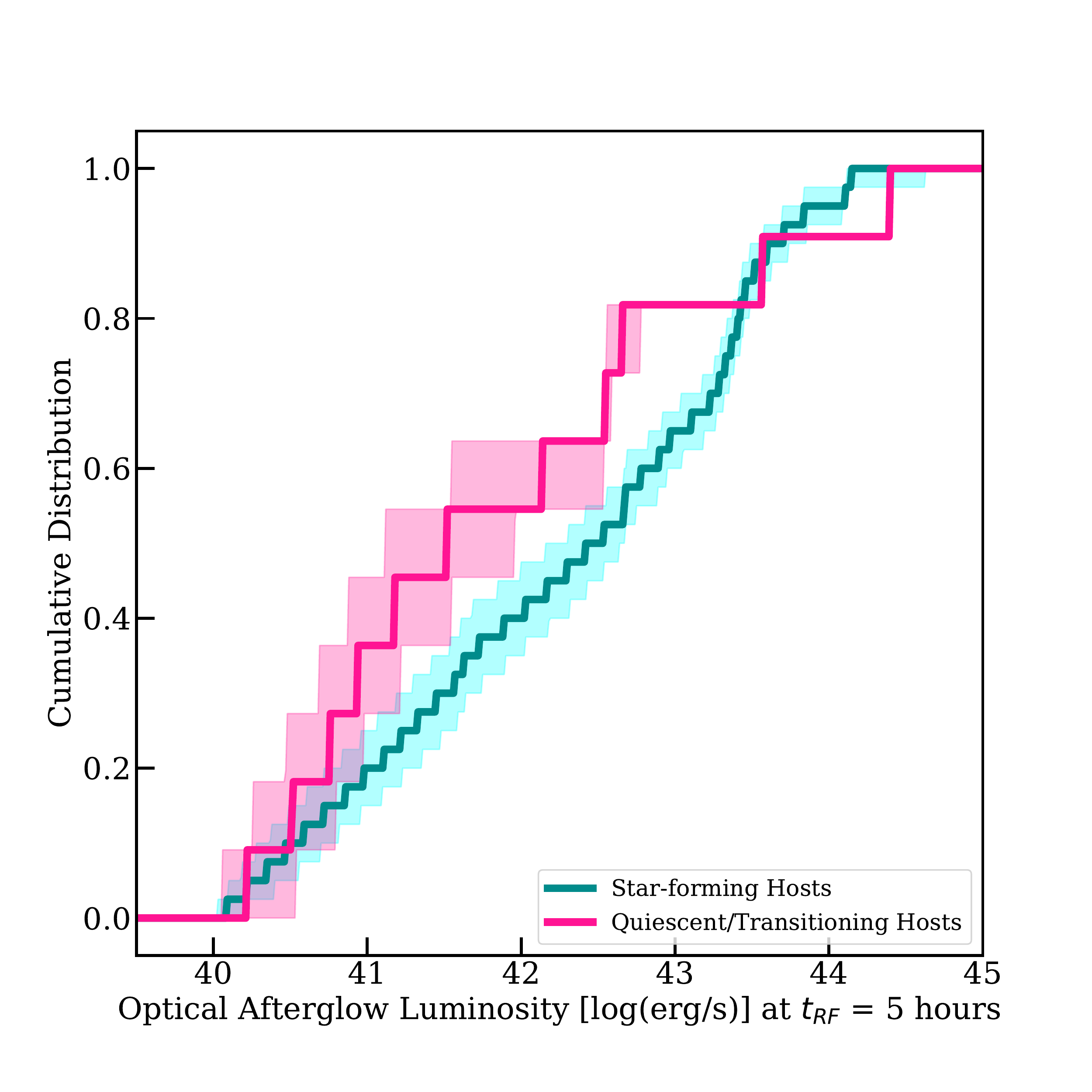}
\vspace{-0.2in}
\caption{The optical afterglow luminosities of short GRBs at a common rest-frame time of 5 hr in the host frame. While the optical afterglows appear to be fainter in quiescent and transitioning hosts (blue), we find no statistically significant difference when compared to those in star-forming (blue) hosts, implying that luminosity and detectability are not strongly dependent on global host properties.}
\label{AG}
\end{figure}

\begin{figure*}[t]
\makebox[\textwidth][c]{\includegraphics[width=1.0\textwidth]{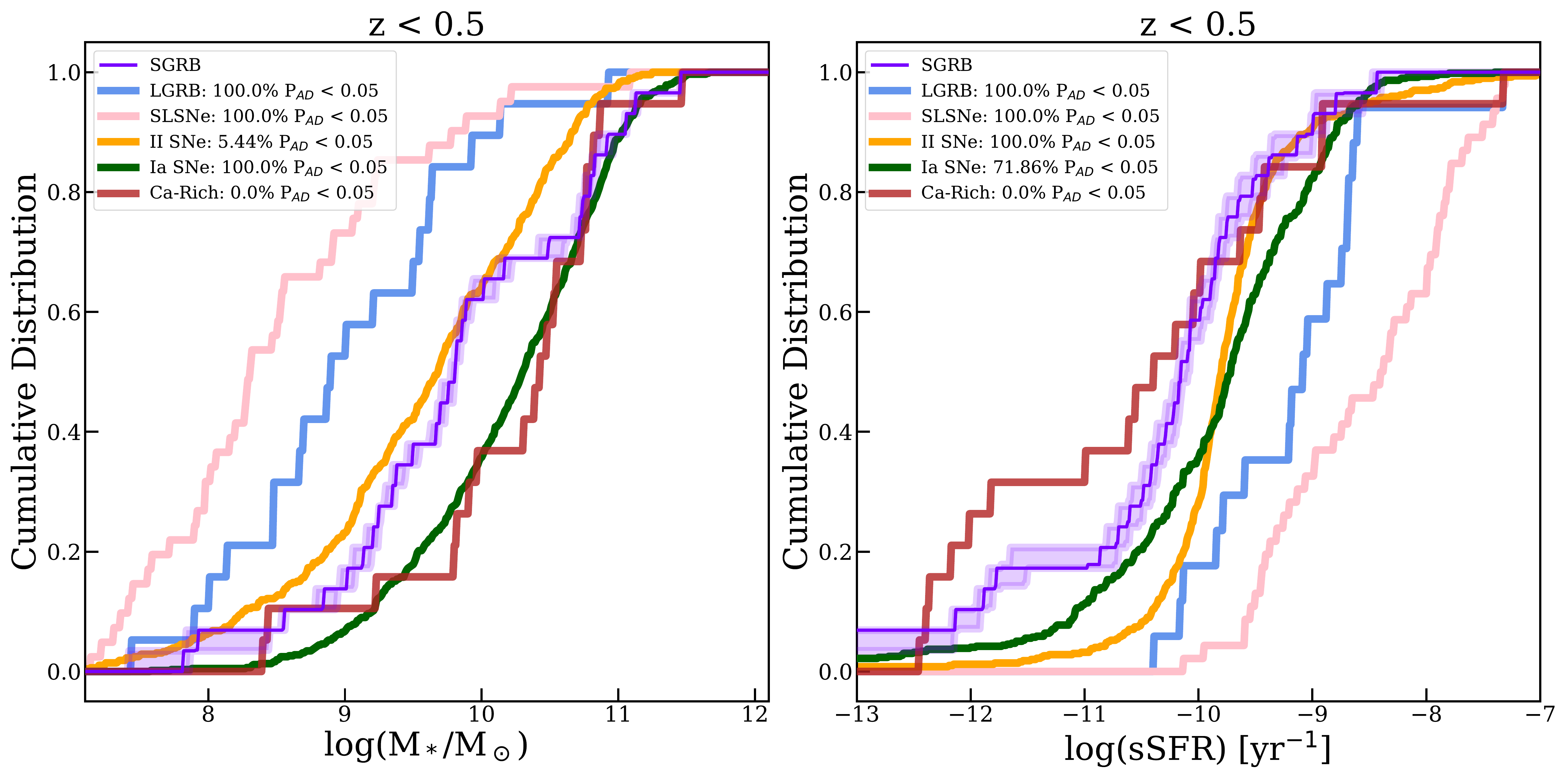}}
\vspace{-0.2in}
\caption{CDFs of log(M$_*$/M$_\odot$) (left) and sSFR (right) for short GRB hosts (purple), Type Ia SNe hosts (green), Type II SNe hosts (orange), Ca-rich transients hosts (red), SLSNe hosts (pink), and long GRB hosts (blue). For the short GRB host CDFs, we draw 5000 random samples from the respective \texttt{Prospector} posterior distribution, and plot the medians of the resulting cumulative distributions as well as the $68\%$ credible region to represent uncertainty (in light purple). We perform Anderson Darling tests between all 5000 short GRB host distributions and the other transient hosts' distributions and show the percentage of tests that reject the null hypothesis that they are derived from the same underlying distribution in the legend. We find Type Ia, Ca-rich transient, and short GRB hosts' sSFRs to be less than those of other transients and find their stellar masses to be higher.}
\label{othertransients}
\end{figure*}

Finally, we explore the circumburst densities of short GRBs along with their host properties. We find circumburst densities for 29 short GRB hosts ($\approx 80\%$ occur in star-forming hosts) in \citet{fbm+15} and \citet{smf+20} and compare them to the hosts' properties. We find no correlation between any host property and the circumburst density, and find that short GRBs in star-forming, transitioning, and quiescent galaxies span the full range of circumburst densities. We also do not see any trend between circumburst densities and the short GRB offsets. As these densities only probe the local environment around the GRB, we infer from this that short GRBs in star-forming and quiescent galaxies both can occur in low-density environments. It may also indicate that we do not have the sensitivity to detect any difference between short GRB densities in either host group.

\section{Comparisons to Transient Environments and Implications for Gravitational Wave Follow-up}
\label{sec:sgrb_trans}
\subsection{Supernovae and Long GRBs}
\label{sec:othertrans}
To provide context for the short GRB host population and the conditions needed to form their progenitors, we now compare their host stellar population properties to those of other well-studied transients. We use the hosts of Type Ia SNe, Ca-rich transients, Type II SNe,  SLSNe, and long GRBs, as these transients originate from a wide variety of progenitor models. We use the stellar population properties in \citet{Pan2021} (Type Ia SNe at $z <$ 0.09), \citet{Wiseman2021} (Type Ia SNe at 0.2 $< z <$ 0.6), \citet{dml+2022} (Ca-rich transients), \citet{Svensson2010, Perley2013, Vergani2015, Wang2014, Niino2017} (long GRBs), and \citet{Schulze2021} (SLSNe and Type II SNe). We caution that these studies use a variety of methods for determining stellar population properties, including applying different SED modeling codes, stellar mass and SFR relations based on colors, and elliptical and spiral galaxy stellar population templates, and thus, the very precise comparisons here are not possible. We choose stellar mass and sSFR as a basis for comparison as they are the most cataloged host stellar population properties for the other transients, and stellar mass is generally fairly impervious to modeling differences. We choose to only compare the stellar masses and sSFRs at $z < 0.5$ for all transients as this is where there exists the most overlap among short GRBs and the comparison samples. This is also the redshift range where a majority of short GRBs lie, thus we have a large sample to compare to the other transient hosts. 

We plot the CDFs of the other transient hosts from their inferred log(M$_*$/M$_\odot$) and sSFRs, and compare to short GRBs in Figure \ref{othertransients}. We initially find that long GRBs and SLSNe, which both are derived from young massive star progenitors, occur in hosts with smaller stellar masses and higher rates of star formation than those of short GRBs. Type Ia SNe, which have older stellar progenitors, appear to share similar host properties to those of short GRBs. We also find that the stellar mass distributions of Ca-rich transients are very similar to those of short GRBs, although only a portion of Ca-rich transients may truly be derived from older stellar progenitors, while the rest are likely from massive stars. To further quantify this, we perform AD tests and can reject the null hypothesis that short GRB hosts are derived from the same stellar mass and sSFR distributions as Type II, long GRB, and SLSNe host galaxies. In particular, this reinforces the distinct nature of short GRB environments from those which bear massive star explosions. In contrast, we cannot reject the null hypothesis that short GRB hosts' sSFRs are derived from the same underlying distributions as those of Type Ia and Ca-rich transients, as well as the stellar masses between short GRBs and Ca-rich transients (Type Ia SNe hosts appear to have slightly higher stellar masses). Since we know that Type Ia supernovae originate from evolved single- or double-degenerate progenitors and have an observed power-law DTD ($p(t) \propto t^{-1}$; \citealt{peters1964, bkv2002, Totani2008, mmb12, Graur2014, bpp2019}), it comes as no surprise that their global host environments are similar to those of short GRBs. While the progenitors of Ca-rich transients are uncertain, with models ranging from white dwarf origins in globular clusters \citep{Shen2019} to core collapse from a stripped-envelope low-mass massive star \citep{de+2021}, it is clear that their progenitors do not select starkly different environments from short GRBs. We also note that the Ca-rich sample only extends to $z \sim 0.02$, over which galaxies in general have lower inferred sSFRs and higher stellar masses; thus, our observations could also be a product of selection effects of the current Ca-rich host sample.

\subsection{Comparison to GW170817 and Implications for Gravitational Wave Follow-Up}
\label{gwfollowup}
The joint detection of GW170817 with the short GRB\,170817A provides direct evidence on the origins of some short GRBs. Thus, the host environments of short GRBs may provide information on the environmental properties of local, GW-detected mergers. In this work, we have compared the environment of GW170817, the only GW-detected BNS merger, with a host galaxy (NGC4993) and summarize our findings here. With an age of $\approx 10.41$~Gyr and an sSFR $\approx -12.34$~yr$^{-1}$, NGC4993 is older and has a lower sSFR than all galaxies in the sample. It is fairly massive, at the $\sim83.5$\% percentile for stellar mass although not the most massive host. The fact that NGC4993 is unique galaxy to host NS mergers is also supported by simulations of BNS merger environments. Indeed, \citet{mlt+21} found NGC4993 to have one of the lowest SFRs compared to its sample of simulated BNS and NSBH mergers at $z<0.4$, and instead finds much higher probabilities that star-forming galaxies will host these mergers. While NGC4993 might be unusual compared to the hosts at higher redshifts, it may not be that extraordinary of a host given galaxy evolution, as the low redshift galaxy population has a more significant fraction of quiescent galaxies (see Figure \ref{SFMS}). Thus, as GW detectors are still constrained to local universe BNS and NSBH events ($z < 0.04$ for BNS mergers and $< 0.08$ for NSBH mergers in LIGO/Virgo O4; \citealt{ALIGO_det}) and given that short GRB hosts trace the SFMS, we might expect the majority of NS merger hosts will be star-forming, but with a more prominent fraction of quiescent hosts than at higher redshifts. If all local mergers are derived from NGC4993-like galaxies, this would indicate that the population may solely be tracing stellar mass, in which case we could infer that local mergers exclusively represent a long-delay time population. On the other hand, GRB 211211A has a host with properties that contrast greatly from those of NGC4993. Thus, if this event is truly derived from an NS merger, this would indicate that mergers in the local universe also trace a combination of stellar mass and star formation, similar to the cosmological merger population. 

One of the major challenges of identifying EM counterparts to GW events are the large localization regions that need to be searched. Many optical efforts have turned to galaxy-targeted searches, which search around potential host galaxies for counterparts. At present, such follow-up searches generally rank galaxies by either their $B$-band luminosities, which is correlated with star formation, or less frequently their $K$-band luminosity, which traces the stellar mass \citep{white2011, bbf+17, pht+17, dalya+2018, eby+2020, kfb+2022}. The results from our catalog, if representative of the local universe population, suggest that ranked-based searches should not singularly depend on either the $B$- or $K$-band luminosities. Our results suggest that follow-up searches that invoke these observables should involve a combination of these luminosities, as the most likely host for a BNS event will not always be the most massive nor the most star-forming galaxy. Our study also demonstrates that short GRBs span a diverse range of environments, and thus, ranked-based searches will not always be successful in identifying the most likely hosts.

\section{Sample Biases}
\label{sec:biases}
Finally, we discuss potential biases in the full sample and sub-samples of short GRB host galaxies. \citet{BRIGHT-I} discussed selection effects and the use of the probability of chance coincidence ($P_\textrm{cc}$) method to associate short GRBs to their hosts. There, we used inclusive criteria for host identification and found more high-redshift (e.g., $z>1$) and low-luminosity hosts than previously reported. In this paper, we further find that there are almost no hosts with stellar masses of $\lesssim 10^8$ M$_\odot$ in our sample (except for GRB 060614 210726A's hosts), consistent with the lack of low-luminosity hosts despite our survey being sensitive enough \citep{BRIGHT-I}. This may imply that short GRBs do not often occur in low mass or dwarf galaxies because they produce less of the total stars and BNS systems in the universe, or that it is difficult to associate short GRBs to such hosts below the mass completeness limit. As a test, we create a \texttt{Prospector} model SED at $z=0.6$ (the median redshift of host sample), M$_F= 10^8$M$_\odot$, and the medians of the rest of the stellar population properties. We find that the host would have an $i$-band magnitude of $\approx 26$ mag AB, at the limit of ground-based telescopes. If a progenitor experienced a significant kick from such a galaxy, and therefore exhibits even a moderate offset, the $P_\textrm{cc}$ method would not generally favor such a host.

We also note that there are 21 short GRBs in the \citet{BRIGHT-I} sample that either do not have confirmed hosts (6 short GRBs) or do not have enough data for a \texttt{Prospector} fit (17 short GRBs), despite the fact that follow-up observations for these bursts were not limited by any significant observing constraint. Many of the latter hosts are simply too faint to be detected in multiple photometric bands with a single {\it HST} detection, and therefore have low luminosities and/or high redshifts (and likely low masses). If we assume an extreme case that all of these hosts exist at $z \geq 1.0$, the redshift median of the short GRB sample would increase to $z \approx 0.9$. Further, if all were low mass ($\leq 10^{8.5}$ M$_\odot$), the mass median would decrease to log(M/M$_\odot$) $\approx 9.4$. Finally, if all these hosts were quiescent galaxies, the star-forming fraction would decrease to $\approx 64 \%$, which is more consistent with the results from \citet{mlt+21}. However, given that we have evidence for disk morphologies for several of these hosts (cf., \citealt{fb13}), it is unlikely that all of them are quiescent.

We next explore how results may change if we only analyze the Gold Sample hosts. As shown in Figure \ref{gsb}, there are no significant differences in stellar population properties between the Gold, Silver, and Bronze Samples, except potentially in redshift. Figure 8 in \citet{BRIGHT-I} also shows that $P_\textrm{cc}$ increases (poorer association) with increasing redshift, thus removing the Silver and Bronze samples from any host analysis might bias our results slightly in favor of more low-redshift, brighter hosts. Since there are several redshift-dependent parameters explored in this paper, including stellar mass, SFR, and age, only examining the Gold Sample might decrease the significance of fast-merging short GRB progenitors occurring in young, star-forming galaxies. 

A more significant bias in understanding short GRB hosts occurs if we only explore the spectroscopic redshift as opposed to the full sample. We find a significant difference in the redshift ranges (no significant changes between other stellar population properties) between the spectroscopic and photometric redshift samples, as the photometric redshift sample contains many $z \gtrsim 1.0$ high-redshift hosts. By not including the photometric redshift sample, we are undoubtedly excluding short GRB progenitors that formed and merged in the early universe, which has significant implications for their DTD. As short GRBs contribute to the $r$-process element nucleosynthesis, this may also affect our understanding of chemical enrichment of the universe. Therefore, in the absence of spectroscopic confirmation with, e.g., the {\it James Webb Space Telescope}, the inclusion of the photometric redshift sample will play a crucial role in understanding the environments and progenitors of short GRBs.

\section{Conclusions}
\label{sec:conclusions}
In this paper, we modeled 69 short GRB host galaxies with \texttt{Prospector} to determine their redshifts and stellar population properties, including stellar mass, mass-weighted age, (specific) SFR, metallicity, and dust attenuation. We used the host galaxies described in \citet{BRIGHT-I}, which are divided into three sub-samples based on host association confidence: Gold (best associated), Silver, and Bronze Samples. For this work, we only model host galaxies when there are at least three host detections in different photometric filters. As the largest catalog of short GRB host properties to date, this sample has tremendous implications for their progenitor systems. We summarize our main conclusions below:
\begin{itemize}
    \item Short GRBs span a spectroscopically confirmed redshift range of $0.1 \leq z \leq 2.211$, with a median of $z \approx 0.47$. When including the short GRBs with photometric redshifts, which fills in the ``redshift desert" at $z \gtrsim 1.0$, the median increases to $z \approx 0.64$. The inclusion of the photometric redshift sample is important in quantifying the number of high-redshift, fast-merging progenitors. The Silver and Bronze Sample hosts span higher redshifts than the Gold Sample hosts.
    \item The full short GRB host sample has $\log(M_*/M_\odot) = 9.69^{+0.75}_{-0.65}$, SFR $= 1.44^{+9.37}_{-1.35}$ M$_\odot$yr$^{-1}$, $t_m = 0.80^{+2.71}_{-0.53}$ Gyr, $\log(Z_*/Z_\odot) = -0.38^{+0.44}_{-0.42}$, and $A_V =  0.43^{+0.85}_{-0.36}$ mag. The Gold, Silver, and Bronze Samples all have stellar population properties that are consistent with the full sample.
    \item Approximately $84 \%$ of short GRB hosts are classified as star-forming galaxies. Star-forming hosts dominate the host population at $z > 0.25$, and the star-forming fraction roughly follows that of the observed field galaxy population at these redshifts. At $z < 0.25$, we find a growing presence of quiescent and transitioning hosts. The frequency of short GRB hosts across redshift are fully consistent with the field galaxy population.
    \item We find that across all redshifts, short GRBs have SFRs consistent with their stellar masses and populate the range of the SFMS. However, they do not trace the stellar mass distribution of field galaxies alone. This implies that short GRBs trace a combination of recent star formation and stellar mass, but their rate is not singularly dependent on either property.
    \item Stellar population ages of short GRBs span a wide range ($\approx 0.1-9$~Gyr), implying a spread in progenitor delay times. The majority of ages are $\leq 1$~Gyr. At $z>1$, their ages are consistently only $\lesssim 20\%$ of a Hubble time at their respective redshifts, indicating the dominance of fast-merging systems at these epochs. As redshift decreases, the population of hosts expands to include those with ages closer to the Hubble times at those redshifts. This indicates that systems at low redshifts likely formed closer to the epoch of peak star formation in the universe and have long delay times. This also suggests a decreasing NS binary formation efficiency from high to low redshifts.
    \item Short GRB progenitors do not have a strong preference for stellar metallicity, but instead their hosts have a spread in metallicities that matches the field galaxy population. As the amount of metal content is inversely correlated to the amount of star formation, this shows that there are some short GRB progenitors directly linked to recent star formation, as well as progenitors very removed from it. 
    \item We find that most short GRBs in quiescent hosts appear to have shorter $T_{90}$ values and lower fluences than those in star-forming hosts. We also see evidence for larger migrations overall from short GRBs in non star forming hosts than those in star-forming hosts. We find no correlation between host properties and short GRB  optical afterglow detection, optical afterglow luminosity, and circumburst density. We also find that the population of bursts with extended emission is drawn from the same underlying distribution in terms of all tested properties, indicative of the same or similar progenitors for both populations.
    \item Short GRB hosts have lower sSFRs and higher stellar masses than those of SLSNe, and long GRBs and Type II SNe, while they have similar such properties to the hosts of Type Ia SNe and Ca-rich transient hosts.
    \item Compared to the full sample of short GRB hosts, NGC4993 (host of GW170817/GRB\,170817A) is older than any other host and has a lower sSFR. It is fairly massive, at the $84 \%$ level for stellar mass. While quiescent, massive galaxies such as NGC4993 are more prevalent in the local horizon probed by GW detectors, the short GRB host population exhibits broader diversity even at low redshifts. Indeed, if the short GRB host population is representative of GW-detected merger hosts, then a combination of $B$- and $K$-band luminosities should be used in targeting potential host galaxies.
\end{itemize}

\noindent Overall, our work demonstrates that short GRB progenitors can form in a variety of environments, across a wide range of redshifts and star formation histories. We find evidence for both a short delay time population, which appears in young and actively star-forming stellar populations and are the majority of observed events, and a long delay time population, which becomes more prevalent at lower redshifts in older and quiescent galaxies. Our work paves the way for quantitative constraints on the true DTD by convolving the \texttt{Prospector} derived SFHs with the stellar population ages and redshifts (see \citealt{Zevin+DTD}).

Through modeling with \texttt{Prospector}, we have significantly augmented the known ranges of redshifts and stellar population properties probed by short GRBs. As NS mergers that lead to short GRBs are known to produce some (if not most or all) of the $r$-process elements in the universe, understanding their true redshift distribution allows us to better comprehend when in the universe's history these elements were created and the contribution from mergers. Furthermore, we can use this host stellar population catalog to better associate GW-detected BNS and NSBH mergers to their host galaxies, especially as the horizon of ground-based GW detectors increases. In light of the fact that short GRBs trace a combination of stellar mass and recent star formation, future galaxy-targeted searches for EM counterparts could be optimized to account for this information.  At present, short GRBs are detected more frequently and out to farther cosmological distances than GW-detected mergers. Thus, this catalog provides a legacy sample for comparison that will not be matched by GW detections for many years to come. Furthermore, this catalog of host properties holds the promise to quantify specific NS binary traits, such as their natal kicks, orbital separations, and delay times.

We have provided a comprehensive catalog of short GRB host properties based on nearly two decades of discoveries. However, there remain bursts with no confirmed or very faint hosts that need characterization. In addition, deriving the true spectroscopic redshifts for the photometric sample will solidify the redshift distribution especially at $z>1$. The \textit{James Webb Space Telescope} will be pivotal in both endeavors, providing unprecedented high-quality spectra for inference on their stellar populations.

\section*{Acknowledgements}
We thank the anonymous referee for valuable insight and suggestions. A.E.N. acknowledges support from the Henry Luce Foundation through a Graduate Fellowship in Physics and Astronomy. The Fong Group at Northwestern acknowledges support by the National Science Foundation under grant Nos. AST-1814782, AST-1909358 and CAREER grant No. AST-2047919. W.F. gratefully acknowledges support by the David and Lucile Packard Foundation, the Alfred P. Sloan Foundation, and the Research Corporation for Science Advancement through Cottrell Scholar Award 28284.
Support for M.Z. was provided by NASA through the NASA Hubble Fellowship grant HST-HF2-51474.001-A awarded by the Space Telescope Science Institute, which is operated by the Association of Universities for Research in Astronomy, Incorporated, under NASA contract NAS5-26555. Y.D. is supported by the National Science Foundation Graduate Research Fellowship under Grant No. DGE-1842165. AL is supported by the European Research Council (ERC) under the European Union's Horizon 2020 research and innovation programme (grant agreement No. 725246). 

This work is based on observations taken by the 3D-HST Treasury Program (GO 12177 and 12328) with the NASA/ESA HST, which is operated by the Association of Universities for Research in Astronomy, Inc., under NASA contract NAS5-26555. 

This research has made use of the HST-COSMOS database, operated at CeSAM/LAM, Marseille, France.

This research was supported in part through the computational resources and staff contributions provided for the Quest high performance computing facility at Northwestern University which is jointly supported by the Office of the Provost, the Office for Research, and Northwestern University Information Technology.
 
\vspace{5mm}
\facilities{Swift (XRT)}
\vspace{5mm}
\software{\texttt{Prospector} \citep{Leja_2017}, \texttt{Python-fsps} \citep{FSPS_2009, FSPS_2010}, \texttt{Dynesty} \citep{Dynesty}}

\pagebreak
\startlongtable
\begin{deluxetable*}{l|ccccccccc}
\tabletypesize{\footnotesize}
\tablecolumns{10}
\tablewidth{0pc}
\tablecaption{Short GRB Host Galaxy Stellar Population Properties
\label{tab:prospectres}}
\tablehead{
\colhead{GRB}	 &
\colhead{$z$} &
\colhead{Galaxy Type} &
\colhead{$t_m$ [Gyr]} &
\colhead{log(M$_*$/M$_\odot$)} &
\colhead{SFR [M$_\odot$/yr]} &
\colhead{log(sSFR) [yr$^{-1}$]} &
\colhead{log(Z$_*$/Z$_\odot$)} &
\colhead{A$_V$ [mag]} &
\colhead{Fit Type} 
}
\startdata
\multicolumn{10}{c}{GOLD SAMPLE} \\
\hline
050509B & $0.225$ & Q & $8.84^{+0.02}_{-0.04}$ & $11.46^{+0.0}_{-0.0}$ & $0.21^{+0.01}_{-0.01}$ & $-12.14^{+0.02}_{-0.02}$ & $-0.29^{+0.01}_{-0.01}$ & $0.0^{+0.0}_{-0.0}$ & S \\
050709 & $0.161$ & SF & $0.57^{+0.0}_{-0.0}$ & $8.55^{+0.01}_{-0.01}$ & $0.02^{+0.0}_{-0.0}$ & $-10.17^{+0.01}_{-0.01}$ & $-0.25^{+0.01}_{-0.01}$ & $0.02^{+0.03}_{-0.02}$ & S \\
050724 & $0.257$ & Q & $8.2^{+0.2}_{-0.33}$ & $11.05^{+0.01}_{-0.01}$ & $0.15^{+0.01}_{-0.01}$ & $-11.89^{+0.04}_{-0.04}$ & $-0.63^{+0.04}_{-0.04}$ & $0.76^{+0.04}_{-0.04}$ & S \\
051221A & $0.546$ & SF & $0.49^{+0.07}_{-0.06}$ & $9.31^{+0.03}_{-0.03}$ & $0.71^{+0.11}_{-0.09}$ & $-9.46^{+0.06}_{-0.06}$ & $-0.97^{+0.04}_{-0.02}$ & $0.35^{+0.09}_{-0.1}$ & S \\
060614 & $0.125$ & SF & $0.76^{+0.2}_{-0.14}$ & $7.85^{+0.04}_{-0.03}$ & $0.0^{+0.0}_{-0.0}$ & $-10.17^{+0.05}_{-0.06}$ & $-0.82^{+0.14}_{-0.11}$ & $0.1^{+0.12}_{-0.06}$ & S \\
060801 & $1.13$ & SF & $0.13^{+0.07}_{-0.04}$ & $9.12^{+0.09}_{-0.09}$ & $9.19^{+2.79}_{-2.11}$ & $-8.15^{+0.15}_{-0.16}$ & $-0.12^{+0.1}_{-0.12}$ & $0.3^{+0.11}_{-0.1}$ & S \\
061006 & $0.461$ & SF & $4.27^{+1.01}_{-1.18}$ & $9.37^{+0.07}_{-0.08}$ & $0.1^{+0.07}_{-0.03}$ & $-10.37^{+0.27}_{-0.19}$ & $-0.84^{+0.19}_{-0.11}$ & $0.24^{+0.24}_{-0.15}$ & S \\
061210 & $0.41$ & SF & $0.66^{+0.13}_{-0.11}$ & $9.49^{+0.04}_{-0.04}$ & $0.19^{+0.26}_{-0.12}$ & $-10.21^{+0.37}_{-0.43}$ & $-0.03^{+0.12}_{-0.26}$ & $0.45^{+0.24}_{-0.24}$ & P \\
070429B & $0.902$ & SF & $0.43^{+0.06}_{-0.04}$ & $10.44^{+0.02}_{-0.02}$ & $8.65^{+2.07}_{-1.78}$ & $-9.5^{+0.09}_{-0.1}$ & $-0.93^{+0.08}_{-0.05}$ & $2.0^{+0.11}_{-0.11}$ & S \\
070714B & $0.923$ & SF & $1.62^{+0.68}_{-0.55}$ & $9.37^{+0.08}_{-0.08}$ & $1.22^{+0.48}_{-0.44}$ & $-9.29^{+0.23}_{-0.23}$ & $-0.51^{+0.3}_{-0.27}$ & $0.33^{+0.17}_{-0.18}$ & S \\
070724 & $0.457$ & SF & $0.27^{+0.0}_{-0.0}$ & $9.81^{+0.0}_{-0.0}$ & $6.49^{+0.1}_{-0.1}$ & $-8.99^{+0.01}_{-0.01}$ & $-0.75^{+0.0}_{-0.0}$ & $1.25^{+0.02}_{-0.02}$ & S \\
070809 & $0.473$ & T & $0.84^{+0.33}_{-0.13}$ & $10.82^{+0.02}_{-0.06}$ & $0.83^{+1.34}_{-0.71}$ & $-10.9^{+0.41}_{-0.76}$ & $-0.22^{+0.26}_{-0.4}$ & $1.05^{+0.14}_{-0.26}$ & P \\
071227 & $0.381$ & SF & $1.78^{+1.65}_{-0.73}$ & $10.49^{+0.13}_{-0.07}$ & $5.78^{+5.29}_{-3.41}$ & $-9.75^{+0.33}_{-0.43}$ & $-0.16^{+0.21}_{-0.25}$ & $2.18^{+0.3}_{-0.41}$ & P \\
090510 & $0.903$ & SF & $0.45^{+0.03}_{-0.03}$ & $9.75^{+0.01}_{-0.01}$ & $1.26^{+0.24}_{-0.21}$ & $-9.65^{+0.08}_{-0.08}$ & $-0.96^{+0.06}_{-0.03}$ & $0.54^{+0.1}_{-0.1}$ & S \\
100117 & $0.914$ & Q & $3.02^{+1.09}_{-0.79}$ & $10.35^{+0.08}_{-0.08}$ & $0.0^{+0.09}_{-0.0}$ & $< -10.80$ & $-0.45^{+0.2}_{-0.2}$ & $0.16^{+0.12}_{-0.1}$ & S \\
100206A & $0.407$ & SF & $4.58^{+0.15}_{-0.2}$ & $10.72^{+0.02}_{-0.02}$ & $7.64^{+0.94}_{-0.83}$ & $-9.84^{+0.05}_{-0.04}$ & $0.18^{+0.01}_{-0.01}$ & $1.2^{+0.11}_{-0.11}$ & S \\
101224A & $0.454$ & SF & $0.46^{+0.17}_{-0.08}$ & $9.17^{+0.06}_{-0.06}$ & $0.58^{+0.18}_{-0.13}$ & $-9.4^{+0.11}_{-0.12}$ & $-0.8^{+0.3}_{-0.14}$ & $0.2^{+0.16}_{-0.13}$ & S \\
120804A & $1.05^{+0.23*}_{-0.09}$ & SF & $0.35^{+0.74}_{-0.24}$ & $9.81^{+0.33}_{-0.27}$ & $18.79^{+12.62}_{-8.09}$ & $-8.52^{+0.49}_{-0.57}$ & $-0.05^{+0.17}_{-0.45}$ & $2.77^{+0.4}_{-0.72}$ & P \\
121226A & $1.37^{+0.05*}_{-0.06}$ & SF & $0.12^{+1.09}_{-0.07}$ & $9.46^{+0.55}_{-0.21}$ & $24.7^{+14.47}_{-15.04}$ & $-8.05^{+0.38}_{-1.02}$ & $-0.02^{+0.13}_{-0.16}$ & $1.01^{+0.16}_{-0.46}$ & P \\
130603B & $0.357$ & SF & $1.63^{+0.43}_{-0.52}$ & $9.82^{+0.05}_{-0.04}$ & $0.44^{+0.22}_{-0.09}$ & $-10.18^{+0.22}_{-0.13}$ & $-0.4^{+0.44}_{-0.08}$ & $0.29^{+0.16}_{-0.11}$ & S \\
140129B & $0.43$ & SF & $1.65^{+0.18}_{-0.16}$ & $9.03^{+0.05}_{-0.05}$ & $0.06^{+0.01}_{-0.01}$ & $-10.23^{+0.07}_{-0.06}$ & $-0.49^{+0.02}_{-0.01}$ & $0.14^{+0.14}_{-0.09}$ & S \\
140903A & $0.353$ & SF & $4.24^{+0.5}_{-0.45}$ & $10.81^{+0.04}_{-0.04}$ & $2.28^{+0.51}_{-0.42}$ & $-10.45^{+0.08}_{-0.07}$ & $-0.99^{+0.02}_{-0.01}$ & $2.99^{+0.22}_{-0.21}$ & S \\
141212A & $0.596$ & SF & $2.38^{+0.93}_{-0.7}$ & $9.71^{+0.08}_{-0.09}$ & $1.17^{+0.45}_{-0.31}$ & $-9.64^{+0.19}_{-0.16}$ & $-0.75^{+0.22}_{-0.15}$ & $0.43^{+0.2}_{-0.18}$ & S \\
150101B & $0.134$ & Q & $4.88^{+0.47}_{-0.44}$ & $11.13^{+0.02}_{-0.02}$ & $0.22^{+0.02}_{-0.02}$ & $-11.78^{+0.05}_{-0.05}$ & $-0.11^{+0.02}_{-0.02}$ & $0.25^{+0.02}_{-0.02}$ & S \\
150120A & $0.46$ & SF & $2.28^{+1.24}_{-0.87}$ & $10.01^{+0.09}_{-0.1}$ & $2.28^{+0.91}_{-0.68}$ & $-9.66^{+0.22}_{-0.22}$ & $-0.45^{+0.21}_{-0.15}$ & $1.1^{+0.23}_{-0.24}$ & S \\
150728A & $0.461$ & SF & $0.15^{+0.01}_{-0.01}$ & $9.35^{+0.01}_{-0.01}$ & $8.13^{+0.66}_{-0.6}$ & $-8.44^{+0.03}_{-0.03}$ & $-0.49^{+0.04}_{-0.04}$ & $0.86^{+0.07}_{-0.06}$ & S \\
160411A & $0.81^{+0.65*}_{-0.43}$ & SF & $0.67^{+1.72}_{-0.43}$ & $8.88^{+0.51}_{-0.42}$ & $1.1^{+6.05}_{-0.96}$ & $-8.83^{+0.48}_{-0.65}$ & $-0.47^{+0.41}_{-0.36}$ & $0.68^{+0.5}_{-0.31}$ & P \\
160525B & $0.64^{+2.03*}_{-0.14}$ & SF & $0.13^{+0.32}_{-0.09}$ & $8.04^{+1.96}_{-0.43}$ & $1.37^{+23.16}_{-0.85}$ & $-8.1^{+0.42}_{-0.56}$ & $-0.57^{+0.45}_{-0.2}$ & $0.14^{+0.19}_{-0.1}$ & P \\
170428A & $0.453$ & SF & $5.14^{+0.0}_{-0.0}$ & $9.68^{+0.02}_{-0.01}$ & $0.4^{+0.02}_{-0.01}$ & $-10.08^{+0.0}_{-0.0}$ & $0.19^{+0.0}_{-0.0}$ & $0.0^{+0.0}_{-0.0}$ & S \\
170728B & $1.272$ & SF & $0.42^{+0.34}_{-0.14}$ & $9.87^{+0.11}_{-0.1}$ & $10.95^{+5.22}_{-4.14}$ & $-8.84^{+0.22}_{-0.23}$ & $0.0^{+0.11}_{-0.19}$ & $0.42^{+0.18}_{-0.19}$ & S \\
170817$^\dagger$ & $0.01$ & Q & $10.41^{+0.43}_{-0.68}$ & $10.61^{+0.01}_{-0.02}$ & $0.02^{+0.0}_{-0.0}$ & $-12.34^{+0.08}_{-0.12}$ & $-0.13^{+0.02}_{-0.02}$ & $0.03^{+0.0}_{-0.01}$ & P \\
180418A & $1.55^{+0.22*}_{-0.45}$ & SF & $0.56^{+0.77}_{-0.38}$ & $9.83^{+0.4}_{-0.63}$ & $12.93^{+20.7}_{-10.57}$ & $-8.76^{+0.52}_{-0.48}$ & $-0.43^{+0.37}_{-0.35}$ & $1.3^{+0.57}_{-0.52}$ & P \\
180618A & $0.52^{+0.09*}_{-0.11}$ & SF & $0.35^{+0.48}_{-0.23}$ & $8.81^{+0.25}_{-0.4}$ & $1.85^{+1.77}_{-1.1}$ & $-8.54^{+0.44}_{-0.39}$ & $-0.58^{+0.34}_{-0.27}$ & $0.32^{+0.34}_{-0.2}$ & P \\
180727A & $1.95^{+0.49*}_{-0.57}$ & SF & $0.54^{+0.58}_{-0.35}$ & $9.23^{+0.66}_{-0.83}$ & $3.09^{+10.66}_{-2.46}$ & $-8.76^{+0.5}_{-0.42}$ & $-0.45^{+0.41}_{-0.39}$ & $0.68^{+0.66}_{-0.39}$ & P \\
181123B & $1.754$ & SF & $0.63^{+0.53}_{-0.35}$ & $9.9^{+0.16}_{-0.19}$ & $11.68^{+7.45}_{-4.01}$ & $-8.84^{+0.4}_{-0.28}$ & $-0.4^{+0.32}_{-0.31}$ & $0.36^{+0.25}_{-0.18}$ & P \\
200219A & $0.48^{+0.02*}_{-0.02}$ & SF & $3.52^{+0.71}_{-1.22}$ & $10.74^{+0.03}_{-0.06}$ & $9.91^{+1.86}_{-1.26}$ & $-9.75^{+0.13}_{-0.08}$ & $0.18^{+0.01}_{-0.01}$ & $0.96^{+0.07}_{-0.06}$ & P \\
200522A & $0.554$ & SF & $0.58^{+0.02}_{-0.02}$ & $9.66^{+0.01}_{-0.01}$ & $2.23^{+0.06}_{-0.05}$ & $-9.31^{+0.01}_{-0.01}$ & $-0.07^{+0.03}_{-0.03}$ & $0.01^{+0.01}_{-0.0}$ & S \\
200907B & $0.56^{+1.39*}_{-0.32}$ & SF & $0.88^{+2.43}_{-0.65}$ & $9.36^{+0.67}_{-0.68}$ & $1.34^{+31.2}_{-1.31}$ & $-8.97^{+0.63}_{-1.55}$ & $-0.4^{+0.36}_{-0.39}$ & $0.59^{+0.76}_{-0.4}$ & P \\
210323A & $0.733$ & SF & $0.56^{+0.82}_{-0.2}$ & $8.77^{+0.26}_{-0.13}$ & $0.34^{+0.1}_{-0.06}$ & $-9.25^{+0.11}_{-0.18}$ & $-0.81^{+0.23}_{-0.12}$ & $0.07^{+0.09}_{-0.05}$ & S \\
210726A & $0.37^{+0.32*}_{-0.17}$ & SF & $1.04^{+1.98}_{-0.7}$ & $7.84^{+0.53}_{-0.5}$ & $0.06^{+0.27}_{-0.05}$ & $-9.07^{+0.54}_{-0.72}$ & $-0.48^{+0.42}_{-0.34}$ & $0.68^{+0.8}_{-0.44}$ & P \\
211023B & $0.862$ & SF & $1.71^{+0.92}_{-0.81}$ & $9.65^{+0.17}_{-0.2}$ & $1.45^{+1.53}_{-0.62}$ & $-9.45^{+0.29}_{-0.3}$ & $-0.19^{+0.26}_{-0.54}$ & $0.42^{+0.5}_{-0.28}$ & S \\
211211A & $0.076$ & SF & $2.52^{+1.24}_{-0.5}$ & $8.84^{+0.1}_{-0.05}$ & $0.07^{+0.01}_{-0.01}$ & $-9.98^{+0.07}_{-0.11}$ & $-0.69^{+0.09}_{-0.2}$ & $0.05^{+0.04}_{-0.03}$ & S \\
\hline
\multicolumn{10}{c}{SILVER SAMPLE} \\
\hline
051210 & $2.58^{+0.11*}_{-0.17}$ & SF & $0.23^{+0.35}_{-0.16}$ & $10.96^{+0.87}_{-0.79}$ & $445.45^{+2545.67}_{-377.29}$ & $-8.35^{+0.5}_{-0.42}$ & $-0.2^{+0.23}_{-0.31}$ & $1.91^{+1.03}_{-0.91}$ & P \\
070729 & $0.52^{+1.19*}_{-0.28}$ & SF & $0.55^{+1.01}_{-0.38}$ & $8.75^{+0.7}_{-0.66}$ & $0.9^{+11.53}_{-0.8}$ & $-8.74^{+0.52}_{-0.57}$ & $-0.57^{+0.44}_{-0.29}$ & $0.31^{+0.46}_{-0.21}$ & P \\
090515 & $0.403$ & Q & $6.34^{+1.48}_{-1.99}$ & $10.92^{+0.05}_{-0.09}$ & $0.0^{+0.0}_{-0.0}$ & $< -11.64$ & $-0.17^{+0.15}_{-0.15}$ & $0.1^{+0.12}_{-0.07}$ & P \\
100625A & $0.452$ & Q & $3.55^{+0.72}_{-0.91}$ & $9.7^{+0.06}_{-0.07}$ & $0.0^{+0.0}_{-0.0}$ & $< -11.64$ & $-0.53^{+0.19}_{-0.15}$ & $0.1^{+0.1}_{-0.06}$ & S \\
101219A & $0.718$ & SF & $0.25^{+0.03}_{-0.02}$ & $9.39^{+0.05}_{-0.06}$ & $3.14^{+1.18}_{-0.94}$ & $-8.9^{+0.09}_{-0.1}$ & $0.17^{+0.02}_{-0.03}$ & $1.5^{+0.22}_{-0.24}$ & S \\
111117A & $2.211$ & SF & $0.19^{+0.22}_{-0.1}$ & $9.63^{+0.22}_{-0.22}$ & $22.14^{+9.13}_{-6.98}$ & $-8.27^{+0.31}_{-0.36}$ & $-0.25^{+0.36}_{-0.6}$ & $0.3^{+0.1}_{-0.11}$ & P \\
120305A & $0.225$ & T & $2.11^{+0.59}_{-1.03}$ & $9.17^{+0.07}_{-0.17}$ & $0.03^{+0.0}_{-0.0}$ & $-10.66^{+0.16}_{-0.1}$ & $-0.98^{+0.03}_{-0.01}$ & $0.02^{+0.02}_{-0.01}$ & S \\
130515A & $0.8^{+0.01*}_{-0.01}$ & T & $0.78^{+0.04}_{-0.07}$ & $10.28^{+0.02}_{-0.02}$ & $0.26^{+0.23}_{-0.09}$ & $-10.87^{+0.27}_{-0.17}$ & $-0.69^{+0.18}_{-0.05}$ & $0.14^{+0.09}_{-0.08}$ & P \\
130822A & $0.154$ & T & $2.16^{+0.45}_{-0.42}$ & $10.16^{+0.05}_{-0.05}$ & $0.3^{+0.05}_{-0.04}$ & $-10.69^{+0.08}_{-0.07}$ & $-0.39^{+0.24}_{-0.11}$ & $0.21^{+0.13}_{-0.12}$ & S \\
140930B & $1.465$ & SF & $0.58^{+0.61}_{-0.4}$ & $9.45^{+0.2}_{-0.28}$ & $5.14^{+4.26}_{-1.64}$ & $-8.75^{+0.53}_{-0.34}$ & $-0.09^{+0.2}_{-0.45}$ & $0.28^{+0.23}_{-0.17}$ & P \\
150831A & $1.18$ & SF & $0.51^{+0.57}_{-0.26}$ & $9.49^{+0.17}_{-0.16}$ & $5.99^{+3.62}_{-2.43}$ & $-8.72^{+0.31}_{-0.35}$ & $-0.4^{+0.38}_{-0.38}$ & $1.02^{+0.3}_{-0.27}$ & S \\
151229A & $0.63^{+0.47*}_{-0.34}$ & SF & $1.79^{+2.61}_{-1.45}$ & $8.79^{+0.35}_{-0.65}$ & $0.22^{+1.78}_{-0.22}$ & $-9.33^{+0.79}_{-2.07}$ & $-0.48^{+0.4}_{-0.33}$ & $0.87^{+0.85}_{-0.52}$ & P \\
160303A & $1.01^{+0.18*}_{-0.23}$ & SF & $1.05^{+0.9}_{-0.51}$ & $9.51^{+0.14}_{-0.16}$ & $2.37^{+2.3}_{-1.76}$ & $-9.13^{+0.37}_{-0.66}$ & $-0.35^{+0.33}_{-0.39}$ & $1.62^{+0.61}_{-0.46}$ & P \\
160624A & $0.484$ & SF & $1.19^{+0.45}_{-0.42}$ & $9.74^{+0.06}_{-0.07}$ & $1.28^{+1.09}_{-0.44}$ & $-9.64^{+0.3}_{-0.22}$ & $-0.02^{+0.16}_{-0.58}$ & $0.55^{+0.38}_{-0.3}$ & S \\
160821B & $0.162$ & SF & $0.58^{+0.02}_{-0.02}$ & $9.24^{+0.0}_{-0.0}$ & $0.24^{+0.01}_{-0.01}$ & $-9.86^{+0.02}_{-0.02}$ & $0.1^{+0.05}_{-0.05}$ & $0.01^{+0.01}_{-0.01}$ & S \\
161001A & $0.67^{+0.02*}_{-0.02}$ & SF & $0.78^{+0.23}_{-0.17}$ & $9.73^{+0.04}_{-0.05}$ & $0.53^{+0.59}_{-0.31}$ & $-10.02^{+0.33}_{-0.37}$ & $-0.19^{+0.24}_{-0.39}$ & $0.45^{+0.29}_{-0.28}$ & P \\
170127B & $2.21^{+0.22*}_{-0.64}$ & SF & $0.32^{+0.53}_{-0.21}$ & $9.51^{+0.53}_{-0.6}$ & $10.25^{+32.99}_{-8.19}$ & $-8.5^{+0.51}_{-0.46}$ & $-0.49^{+0.36}_{-0.31}$ & $0.58^{+0.62}_{-0.41}$ & P \\
180805B & $0.661$ & SF & $0.5^{+0.1}_{-0.09}$ & $9.34^{+0.04}_{-0.04}$ & $1.51^{+0.36}_{-0.26}$ & $-9.17^{+0.1}_{-0.09}$ & $-0.23^{+0.16}_{-0.16}$ & $0.28^{+0.11}_{-0.11}$ & S \\
191031D & $1.93^{+0.22*}_{-1.44}$ & SF & $0.81^{+1.62}_{-0.48}$ & $10.38^{+0.18}_{-0.86}$ & $35.98^{+34.5}_{-35.56}$ & $-8.94^{+0.43}_{-1.19}$ & $-0.19^{+0.28}_{-0.39}$ & $1.41^{+1.64}_{-0.4}$ & P \\
\hline
\multicolumn{10}{c}{BRONZE SAMPLE} \\
\hline
050813 & $0.719$ & SF & $3.72^{+0.45}_{-0.63}$ & $10.31^{+0.03}_{-0.04}$ & $1.53^{+0.52}_{-0.41}$ & $-10.12^{+0.15}_{-0.16}$ & $0.11^{+0.05}_{-0.09}$ & $0.23^{+0.14}_{-0.12}$ & P \\
080123 & $0.495$ & SF & $0.43^{+0.2}_{-0.12}$ & $9.85^{+0.07}_{-0.06}$ & $9.31^{+3.38}_{-2.89}$ & $-8.89^{+0.18}_{-0.21}$ & $0.11^{+0.03}_{-0.07}$ & $0.94^{+0.12}_{-0.17}$ & P \\
140622A & $0.959$ & SF & $0.66^{+0.06}_{-0.05}$ & $10.17^{+0.02}_{-0.02}$ & $6.0^{+0.99}_{-0.88}$ & $-9.39^{+0.06}_{-0.06}$ & $-0.5^{+0.02}_{-0.02}$ & $0.6^{+0.1}_{-0.09}$ & S \\
160408A & $1.91^{+0.38*}_{-0.53}$ & SF & $0.62^{+0.59}_{-0.41}$ & $9.32^{+0.65}_{-0.71}$ & $3.57^{+9.67}_{-2.53}$ & $-8.8^{+0.49}_{-0.38}$ & $-0.46^{+0.38}_{-0.34}$ & $0.49^{+0.52}_{-0.32}$ & P \\
161104A & $0.793$ & Q & $2.27^{+0.27}_{-0.26}$ & $10.23^{+0.04}_{-0.04}$ & $0.06^{+0.04}_{-0.02}$ & $-11.48^{+0.21}_{-0.19}$ & $0.02^{+0.07}_{-0.05}$ & $0.08^{+0.08}_{-0.05}$ & S \\
170728A & $1.493$ & SF & $0.16^{+0.62}_{-0.11}$ & $10.09^{+0.31}_{-0.27}$ & $80.03^{+57.04}_{-46.66}$ & $-8.19^{+0.49}_{-0.69}$ & $-0.3^{+0.13}_{-0.21}$ & $1.72^{+0.2}_{-0.34}$ & S \\
200411A & $0.83^{+0.17*}_{-0.19}$ & SF & $0.63^{+1.0}_{-0.39}$ & $10.23^{+0.2}_{-0.29}$ & $27.35^{+21.94}_{-15.21}$ & $-8.77^{+0.41}_{-0.45}$ & $-0.19^{+0.24}_{-0.35}$ & $1.51^{+0.55}_{-0.41}$ & P \\
201221D & $1.055$ & SF & $0.27^{+0.01}_{-0.01}$ & $9.36^{+0.03}_{-0.03}$ & $2.36^{+0.29}_{-0.26}$ & $-8.98^{+0.05}_{-0.05}$ & $-0.5^{+0.02}_{-0.02}$ & $0.17^{+0.06}_{-0.07}$ & S \\
210919A & $0.242$ & SF & $1.62^{+0.14}_{-0.14}$ & $9.87^{+0.03}_{-0.03}$ & $0.3^{+0.03}_{-0.03}$ & $-10.39^{+0.04}_{-0.04}$ & $-0.49^{+0.08}_{-0.05}$ & $0.8^{+0.09}_{-0.09}$ & S \\
\enddata
\tablecomments{We present the medians and 68\% credible region, or $99.7$th percentile upper limits of the \texttt{Prospector}-derived stellar population properties for all 69 short GRB hosts in our sample, and include the host of GW/GRB 170817 for reference. All values, except redshift when it is known, are set free in the fits. Spectroscopic redshifts are detailed in \citet{BRIGHT-I}. Hosts with spectra used in the \texttt{Prospector} fits have fit type ``S", whereas hosts that only have photometry available for fits have fit type ``P". Gold Sample hosts have the lowest $P_{cc}$ and Bronze Sample hosts have the highest. \\
\footnotesize{$^*$ Photometric redshifts determined through \texttt{Prospector}} \\
\footnotesize{$^\dagger$ Included as a point of comparison, but not included in the short GRB host catalog.}} 
\end{deluxetable*}

\startlongtable
\begin{deluxetable*}{l|ccccccc}
\tabletypesize{\footnotesize}
\tablecolumns{10}
\tablewidth{0pc}
\tablecaption{Stellar Population Properties for Short GRB Host Sub-samples
\label{tab:allgroup}}
\tablehead{
\colhead{Sample} &
\colhead{$z$} &
\colhead{$t_m$ [Gyr]} &
\colhead{log(M$_*$/M$_\odot$)} &
\colhead{SFR [M$_\odot$/yr]} &
\colhead{log(sSFR) [yr$^{-1}$]} &
\colhead{log(Z$_*$/Z$_\odot$)} &
\colhead{A$_V$ [mag]}}
\startdata
 ALL & $0.64^{+0.83}_{-0.32}$ & $0.8^{+2.71}_{-0.53}$ & $9.69^{+0.75}_{-0.65}$ & $1.44^{+9.37}_{-1.35}$ & $-9.47^{+0.92}_{-1.16}$ & $-0.38^{+0.44}_{-0.42}$ & $0.43^{+0.85}_{-0.36}$ \\
 \hline
 GOLD & $0.47^{+0.66}_{-0.22}$ & $0.91^{+3.33}_{-0.63}$ & $9.66^{+1.04}_{-0.81}$ & $1.19^{+8.13}_{-1.1}$ & $-9.57^{+0.97}_{-0.87}$ & $-0.45^{+0.46}_{-0.42}$ & $0.43^{+0.81}_{-0.37}$ \\
 SILVER & $0.7^{+1.38}_{-0.48}$ & $0.74^{+1.77}_{-0.49}$ & $9.6^{+0.67}_{-0.39}$ & $1.28^{+14.11}_{-1.25}$ & $-9.28^{+0.79}_{-1.57}$ & $-0.29^{+0.38}_{-0.45}$ & $0.35^{+1.04}_{-0.29}$ \\ 
 BRONZE & $0.79^{+0.7}_{-0.3}$ & $0.69^{+1.65}_{-0.43}$ & $10.02^{+0.28}_{-0.63}$ & $3.22^{+24.61}_{-2.93}$ & $-9.16^{+0.57}_{-1.23}$ & $-0.3^{+0.4}_{-0.21}$ & $0.66^{+0.77}_{-0.53}$ \\ \hline
 SPECTROSCOPIC REDSHIFT & $0.47^{+0.58}_{-0.25}$ & $1.0^{+3.11}_{-0.67}$ & $9.73^{+0.74}_{-0.51}$ & $1.08^{+6.67}_{-1.02}$ & $-9.76^{+0.84}_{-1.05}$ & $-0.4^{+0.46}_{-0.44}$ & $0.32^{+0.81}_{-0.28}$ \\
 PHOTOMETRIC REDSHIFT & $0.93^{+1.16}_{-0.46}$ & $0.66^{+1.16}_{-0.5}$ & $9.56^{+0.76}_{-1.04}$ & $3.74^{+28.04}_{-3.53}$ & $-8.86^{+0.68}_{-0.96}$ & $-0.34^{+0.39}_{-0.4}$ & $0.85^{+0.91}_{-0.65}$ \\ \hline
STAR-FORMING & $0.66^{+0.8}_{-0.28}$ & $0.63^{+1.86}_{-0.38}$ & $9.6^{+0.6}_{-0.65}$ & $2.22^{+10.82}_{-1.97}$ & $-9.3^{+0.84}_{-0.8}$ & $-0.38^{+0.47}_{-0.43}$ & $0.53^{+0.9}_{-0.42}$ \\
QUIESCENT/TRANSITIONING & $0.4^{+0.4}_{-0.25}$ & $2.87^{+5.22}_{-1.96}$ & $10.35^{+0.76}_{-0.61}$ & $0.14^{+0.14}_{-0.14}$ & $-11.76^{+1.06}_{-2.07}$ & $-0.35^{+0.28}_{-0.36}$ & $0.15^{+0.58}_{-0.13}$
\enddata
\tablecomments{The median and 68\% credible interval stellar population values for the full sample of short GRB hosts, as well as specific sub-samples of hosts based on $P_{cc}$ association (Gold, Silver, Bronze), redshift determination (spectroscopic or photometric), and host type (star-forming or transitioning and quiescent).} 
\end{deluxetable*}
\bibliography{refs}

\appendix 
\restartappendixnumbering

\section{Photometric Redshift Sample Prospector Fits}
\label{sec:photoz}

\begin{figure}[h!]
\centering
\includegraphics[width=1.0\textwidth]{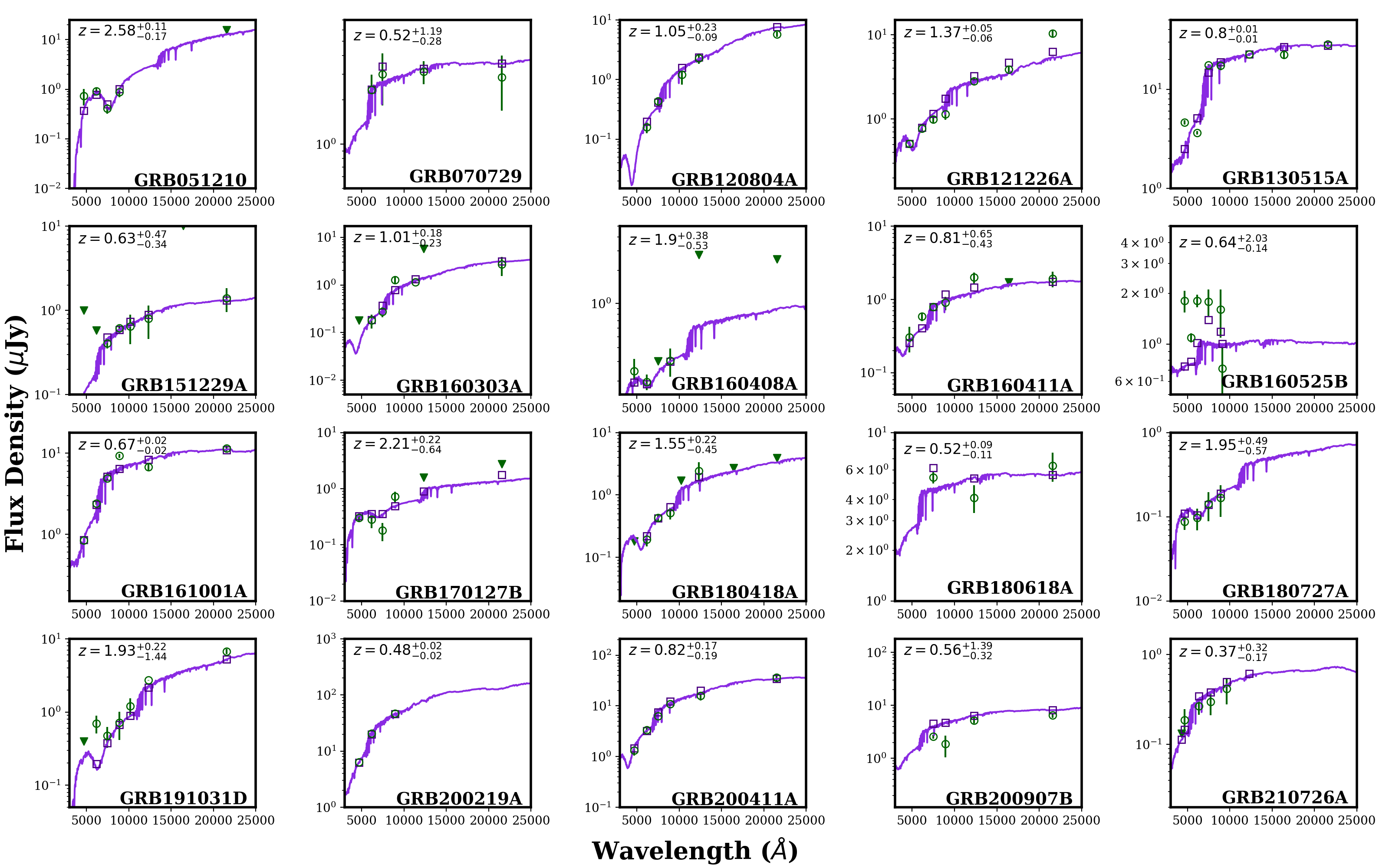}
\caption{Here, we present the \texttt{Prospector}-produced model spectra (purple lines), photometry (purple squares), and observed photometry (green circles) for the 20 short GRBs with no spectroscopic redshift. Upper limits are denoted with green triangles. We find excellent agreement amongst the model and observed SED, suggesting that photometric redshifts are in agreement with the true redshifts of the hosts.}
\label{photoz}
\end{figure}

\section{Literature Comparison of Past GRB Host Properties}
\label{litcomparison}
We compare the short GRB host properties found in this work to previously published properties to explore the differences in stellar population modeling techniques that result in different values for the properties of the hosts. In \citealt{nfd+20}, we found \texttt{Prospector}-derived masses, ages, and redshifts for 19 hosts (GRBs 050509B, 050709, 050724, 050813, 051210, 051221A, 060801, 061006, 061210, 070429B, 070714B, 070724, 070809, 071227, 080123, 090510, 090515, 100117, and 161104A). The \texttt{Prospector} model differs slightly from the one used in this work, as it does not include the mass-metallicity relation, old to young dust attenuation (2:1) ratio, nebular emission line fitting (the gas-phase metallicity and gas ionization parameters), or the spectroscopic noise inflation parameter (when needed). Furthermore, our \texttt{Prospector} fits include the spectra for 12 of the GRB hosts in \citealt{nfd+20} (see Table \ref{tab:prospectres}) which better informs the metallicity, age, and SFR of the host galaxies, whereas \citealt{nfd+20} only included the spectrum for GRB 161104A. We also find stellar population properties for 100206A (SFR, age, mass; \citealt{ber14, csl+2018}), 100625A, 101219A, 111117A, 120804A (SFR's, ages, masses; \citealt{ber14}), 130603B (SFR, mass; \citealt{cpp+13}), 141212A (SFR, mass; \citealt{csl+2018, phc+19}), 150101B (SFR, mass, age; \citealt{fmc+16}), 150120A (SFR, mass; \citealt{csl+2018}), 160821B (SFR, mass; \citealt{tcb+19}), 181123B (SFR, mass, age; \citealt{pfn+20}), and 200522A (SFR, mass, age; \citealt{flr+2021}). We note that \citet{pfn+20} and \citet{flr+2021} also use \texttt{Prospector} modeling, but do not include the mass-metallicity relation as in this work. The rest of the stellar population properties are derived from $B$-band luminosities, spectral line strength, and/or other SED models with different assumed SFHs, including single stellar populations (SSPs) and a declining SFH with bursts of star formation.

We compare the available ages, SFRs, and stellar masses, to the median values in this work. We find large differences in the ages and SFRs of the hosts, which is likely caused by a combination of including spectra in our SED fits, mass-weighted ages, and SED as opposed to emission line SFRs.  Age estimates from SSPs are known to systematically underestimate the true age of the stellar populations, which was the method used to determine the ages for a majority of the previously published values. \citet{Leja2019} also showed that choice of SFH model has the largest effect on the inferred mass-weighted age, thus likely accounting for these differences. Out of the 27 new age estimates, 16 were derived from both the photometry and spectra of the hosts, whereas only 2 of the previously published values used both. Similarly, we performed joint spectroscopic and photometric fits for 8 out of the 12 hosts for new SFR estimates, which had only been done for 1 of the hosts previously. The addition of the available spectra in our fits better constrains the ages and SFRs (and metallicity) of the galaxies than traditional photometry-only based SED fittings, as shown in \citet{jlc+2021} and \citet{Tachella2021}. Indeed, we find that the largest changes in both age and SFR are with the hosts that have a joint spectroscopic and photometric fit. The largest changes in age are with the hosts of GRBs 050509B ($\approx 8$ times larger) and 100206A ($\approx 40$ times larger), and the largest change in SFR is with the host of 160821B ($\approx 6$ times lower). The hosts of GRBs 050509B, 100206A, and 160821B all have spectra with strong spectral lines used in their fits, which likely affected the SFR measurement. We also find that a majority ($\approx 55 \%$) of the new age estimates are higher, and a majority ($\approx 59 \%$) of SFR measurements are lower than previously determined.  In total, the changes in stellar mass, age, and SFR are well understood from comparisons of different stellar population modeling techniques, and the uncertainties from \texttt{Prospector} should properly account for these differences.

We find that the parameter that differs the most between our work and previously literature is the SFR. Thus, we further explore SFR estimates in previously published works when they vary significantly from our inferred SFRs. We note that while we do find differences in the literature of short GRB hosts, we purposely used a method derived in the same way as a comparison field galaxy population to avoid comparing different SFR measurements. We determine that that the majority of discrepancies in SFR are due to using the emission lines exclusively for an estimate, an SED fit with a different model assumption and exclusively using the host photometry, or an SFR inferred from radio data alone. With the exception of GRB\,071227, we find that all SFR estimates determined from emission line strengths are higher than the SED-inferred SFR, as is expected for galaxies \citep{Leja2019}. For example, the hosts of GRBs\,050709 and 101219A have emission-line SFRs $\approx$10 and $\approx$4 times larger, respectively, than the SED-inferred SFRs presented here \citep{ffp+05, cmi+06, pbc+06, fbc+13, ng+2021}. The host of GRB 071227 has a lower reported emission line SFR in \citet{dfp+07} than in this work. There was no available spectrum to use in the \texttt{Prospector} fit for this host; hence, the SFR estimate was more unconstrained, likely leading to this difference. Furthermore, the radio-derived SFR for this host is much larger \citep{ng+2014}, implying there is obscured star formation. As our fit includes NIR data, it likely accounts for this dust, hence why the derived SFR is larger than the emission-line SFR. We find that SFR estimates also substantially change with SED fitting model assumptions, including SFH choice and the use of single versus composite stellar population models (SSP vs. CSP). \citet{pmm+12} and \citet{csl+2018} attempted an exponentially declining SFH for the host of GRB 100206A, a simpler SFH than the delayed-tau model chosen here, which resulted in SFRs 2-3 times larger than in this work. The use of SSPs, which have been shown to less accurately model the true stellar population properties of galaxies and result in younger and less massive galaxies than expected, in \citet{dtr+14} and \citet{phc+19} for the host of GRB 130603B resulted in SFR estimates $\approx$ 10 times greater than those found here. We find that the usage of simpler SFHs, such as exponentially declining SFHs,  and the sole usage of photometry in the SED fits leads to higher inferred SFR estimates. For the host of GRB 100206A, the SFR estimate is 2-3 times larger in \citet{pmm+12} and \citet{csl+2018} and the host of GRB 130603B is $\approx$ 10 times greater in \citet{dtr+14} and \citet{phc+19}.  Finally, as expected for obscured SF probed by radio observations (e.g., \citealt{PerleyPerley2013_RadioSFR, Perley2015, Gatkine2020}), we find the radio-inferred SFRs in the literature always lead to higher values than those from SEDs. This includes the hosts of GRB 071227 \citep{ng+2021}, 100206A \citep{knm+19}, and 120804A (\citealt{bzl+13}; for the latter burst, the lower redshift presented here likely also contributes to a lower inferred SFR).

We find the least variation in stellar mass. As mass is best determined through galaxy photometry and not the spectra, we expect little variation in stellar mass calculations between different SED methods \citep{jlc+2021}. The host of GRB 150101B appears to have the largest change in stellar mass estimate, increasing 1.2 times from the value in \citet{fmc+16}. The previous estimate used a grid-searched SSP model to derive the stellar mass, which has been shown to underestimate the true mass value \citep{Conroy2013SED}. We note that an SSP model for the host of GW/GRB 170817 did resulted in a higher inferred stellar mass estimate in \citet{llt+2017} than that found in our work. Furthermore, this host has a known AGN \citep{xft+2016}, which if not accounted for in the SED fitting, may affect the amount of dust in the infrared, which will lead to errors in the mass inference. We also find larger differences in stellar mass, whether increasing or decreasing, with the galaxies with previously unknown redshifts or current photometric redshifts, where the true SED shape is overall more uncertain. Overall, it does appear that mass estimates from various techniques do result in consistent values.

We finally compare our photometric redshifts to those found in \citet{otd+2022}.  \citet{otd+2022} also used \texttt{Prospector} to determine the redshifts for GRBs\,140129B (we reported a spectroscopic redshift in \citealt{BRIGHT-I}), 170728B (we reported a spectroscopic redshift in \citealt{BRIGHT-I}), 180618A, 151229A, and 200411A. We also identify a different host galaxy for GRB\,191031D due to our deeper NIR imaging in \citet{BRIGHT-I}. They used a restrictive redshift prior range of $0\leq z \leq1.5$ for their analysis. We find this range of sampled values too conservative as \textit{Swift} has detected short GRBs out to much farther confirmed redshifts (e.g. GRB 181123B at $z=1.7$ and GRB 111117A at $z=2.2$; \citealt{skm+18, pfn+20}), suggesting that short GRBs can in principle be detected out to $z \approx 3$. Furthermore, many of our photometric redshift sample hosts have posterior distributions that extend beyond $z = 1.5$; thus, we would not be fully sampling their possible redshifts and other stellar population properties by restricting the maximum to $z = 1.5$.

We emphasize that although we show a majority of the values are consistent with previously published values, it is best practice to compare properties derived from the same technique. Furthermore, as our SED fits include all available photometry, spectroscopy, and redshifts for 69 short GRB hosts, as well as informed priors, our sample represents the largest, uniformly-modeled catalog to date.

\end{document}